\documentclass{aa}  

\usepackage{graphicx}
\usepackage{txfonts}
\usepackage{amsmath}
\usepackage{amsfonts}
\usepackage{natbib}
\usepackage{longtable}
\usepackage{lscape}
\usepackage[usenames, dvipsnames]{xcolor}
\usepackage[normalem]{ulem}
\renewcommand{\arraystretch}{1.2}

\begin{document}

   \title{Searching for debris discs in the 30~Myr open cluster \object{IC 4665}
   }


   \author{N. Miret-Roig\inst{1}
         \and N. Hu\'elamo\inst{2}
         \and H. Bouy\inst{1}
          }

   \institute{Laboratoire d'astrophysique de Bordeaux, Univ. Bordeaux, CNRS, B18N, allée Geoffroy Saint-Hilaire, 33615 Pessac, France.
         \email{nuria.miret-roig@u-bordeaux.fr}
         \and
             Centro de Astrobiología (CSIC-INTA), ESAC Campus, Camino Bajo del Castillo s/n, 28692 Villanueva de la Cañada, Madrid, Spain.
             }

   \date{}

 
  \abstract
   {Debris discs orbiting young stars are key to understand dust evolution and the planetary formation process. We take advantage of a recent membership analysis of the 30~Myr nearby open cluster IC~4665 based on the \textit{Gaia} and DANCe surveys to revisit the disc population of this cluster.}
   {We aim to study the disc population of IC~4665 using \textit{Spitzer} (MIPS and IRAC) and WISE photometry.}
   {We use several colour-colour diagrams with empirical photospheric sequences to detect the sources with an infrared excess. Independently, we also fit the spectral energy distribution (SED) of our debris disc candidates with the Virtual Observatory SED analyser (VOSA) which is capable of automatically detecting infrared excesses and provides effective temperature estimates. }
   {We find six candidates debris disc host-stars (five with MIPS and one with WISE) and two of them are new candidates. We estimate a disc fraction of 24$\pm$10\% for the B--A stars, where our sample is expected to be complete. This is similar to what has been reported in other clusters of similar ages  (Upper Centaurus Lupus, Lower Centaurus Crux, the $\beta$~Pictoris  moving  group,  and  the  Pleiades). For solar type stars we find a disk fraction of 9$\pm$9\%, lower than that observed  in regions with comparable ages.} 
   {Our candidates debris disc host-stars are excellent targets to be studied with ALMA or the future James Webb Space Telescope (JWST).}


   \maketitle
%

\section{Introduction}

Debris discs are the result of collisions between planetesimals and their detection therefore implies that the planet formation process was successful in forming bodies of a few hundreds or a few thousands of kilometres \citep[see e.g. ][for a recent review on debris discs]{Hughes+18}. Stars hosting debris discs are excellent places where to image planets and discs simultaneously, see e.g. Fomalhaut \citep{Kalas+08}, $\beta$~Pictoris \citep{Lagrange+10}, HR8799 \citep{Marois+08}. The reason is that they are optically thin in the infrared, where planets are imaged since they offer the best contrast with respect to the star. On the contrary, protoplanetary discs are optically thick in this wavelengths hindering the observation of planets. In addition, the study of debris discs can give us clues on the composition of exoplanets as well as their orbits and masses \citep[e.g.][]{Hughes+18}.

The first debris disc was discovered around Vega using the \textit{Infrared Astronomical Satellite} (IRAS) by \citet{Aumann+84}. After that, several studies have been devoted to search for debris discs in the solar vicinity \citep[e.g.][]{Moor+06,Rhee+07,Zuckerman+11}. One of the main questions addressed by these studies is the temporal evolution of debris discs, only possible if accurate age measurements are available which is in general not common for isolated stars. The easiest way to tackle this issue is to study debris discs hosted by stars members of a known association or open cluster where the age estimates are much more reliable. In the past decade, a number of studies reported the frequency of infrared (IR) excesses in clusters of different ages (e.g. \citealt{Gorlova+06}, \citealt{Gorlova+07}, \citealt{Siegler+07}). These joint efforts complemented by others on field stars suggested a debris disc fraction decay inversely proportional to the age \citep{Rieke+05}.

IC~4665 is among the sample of young open clusters examined in the literature to search for debris discs. This is a young open cluster with an estimated age of $27.7_{-3.5}^{+4.2}$~Myr \citep{Manzi+08} and an average distance of 350~pc \citep{Miret-Roig+19}. Several works have studied the cluster population \citep{Hogg+55, Prosser+94, deWit+06, Jeffries+09, Lodieu+11}. Recently, \citet{Miret-Roig+19} updated the cluster census using photometric and astrometric information, providing a list of more than 800 highly probable cluster members. \citet{Smith+11} did a search of debris discs in IC~4665 based on \textit{Spitzer} observations. They started from a sample of 75 members and obtained a disc fraction of 27$_{-7}^{+9}$\%. The authors also reported a disc fraction for solar-type stars (F5--K5) of $42_{-13}^{+18}$\% which they claimed to be higher than what had been found in other clusters of similar ages, although compatible within the uncertainties (e.g. \citealt{Gorlova+07} found a F0--F9 fraction of $33_{-09}^{+13}$\% for the 30~Myr NGC~2547 open cluster). 

In this work, we take advantage of \citet{Miret-Roig+19} recent census to revisit the study of debris discs in this cluster. In Section~\ref{sec:data} we describe our sample and dataset which combines photometry from DANCe, WISE and \textit{Spitzer}. In Section~\ref{sec:colour-colour-diagrams} we present our empirical method to detect IR excesses with colour-colour diagrams. In Sect.~\ref{sec:SED} we confirm our candidates by comparing their spectral energy distribution (SED) to models of photospheric emission. In Section~\ref{sec:candidates} we discuss the candidates individually, and in Section~\ref{sec:discussion} we compute the disc fraction and compare it to other studies. Finally, in Section~\ref{sec:conclusions} we present our conclusions.


\section{Data}
\label{sec:data}
We start from a list of 819 candidate members of IC~4665 \citep{Miret-Roig+19} covering a magnitude range of 12.4~mag ($7 < J < 19.4$~mag). This sample has been selected combining photometry and astrometry from the \textit{Gaia} Data Release 2 \citep[DR2, ][]{GaiaCol+16,GaiaCol+18} and the ground-based survey DANCe \citep[Dynamical Analysis of Nearby Clusters, ][]{Bouy+13} in a Bayesian membership algorithm. The resulting sample is expected to be significantly more complete and reliable than the one previously used by \citet{Smith+11} to study debris discs in this cluster. Their sample contained 40 spectroscopic low-mass members from \citet{Jeffries+09}, 33 brighter stars selected with proper motions and $B,V$ photometry from the Tycho-2 catalogue \citep{Hog+00}, and two additional members from \citet{Prosser+94}. The recent \textit{Gaia} DR2 astrometry allows to discard as non-members at a high level of confidence 24 of their 75 targets (32\%), hence motivating a re-analysis of the cluster disc frequency.

\subsection{Photometric database}

\begin{table}[]
    \centering
    \caption{\textit{Spitzer} program IDs used in this study. }
    \label{tab:spitzer_programID}
    \begin{tabular}{ |c|c| } 
     \hline
     Instrument & Program ID  \\ 
      \hline
      IRAC & 13102, 40601, 80072 \\
      MIPS & 3347,40601  \\
     \hline
    \end{tabular}
\end{table}{}

\begin{figure}
    \centering
    \includegraphics[width = \columnwidth]{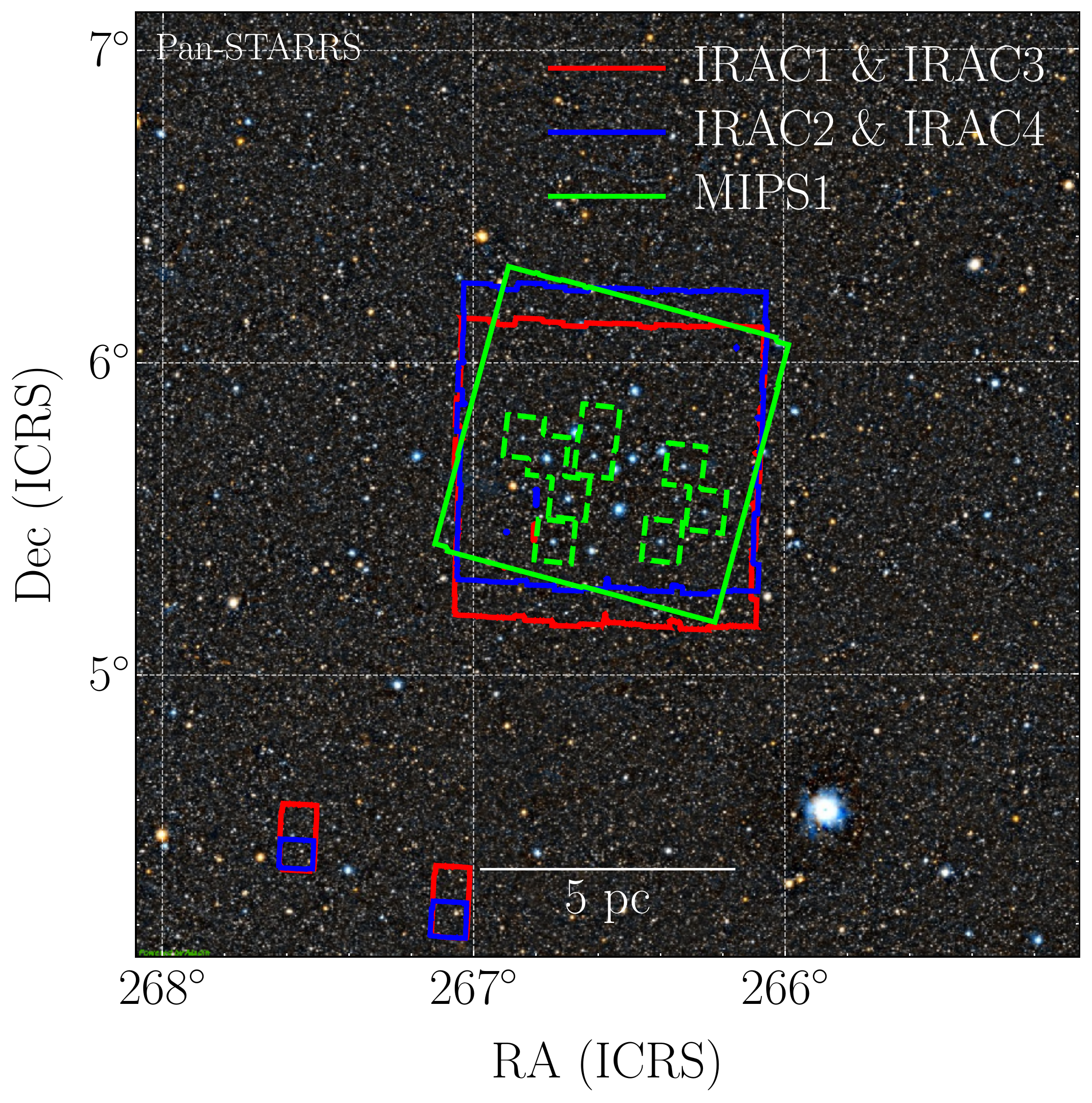}
    \caption{Footprint of the various bands of \textit{Spitzer}. Two MIPS programs cover this cluster namely, program ID 40601 (solid green line) and program ID 3347 (dashed green line). Background image credit: Pan-STARRS. }
    \label{fig:Spitzer_coverage}
\end{figure}{}

\begin{figure*}
    \centering
    \includegraphics[width = 0.95\textwidth]{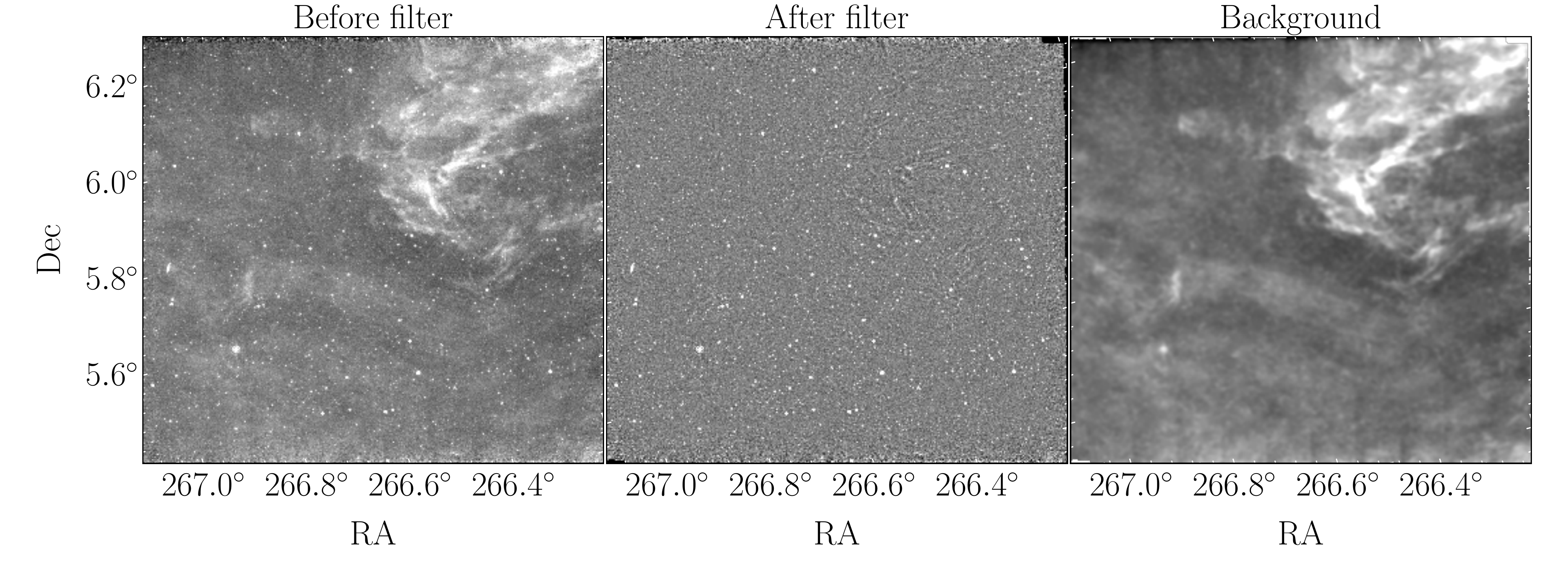}
    \caption{MIPS 24~$\mu$m image before (left) and after (center) the nebulosity filtering. The background computed by the nebulosity filter is shown on the right panel.}
    \label{fig:nebulosity}
\end{figure*}{}

\begin{figure}
    \centering
    \includegraphics[width = \columnwidth]{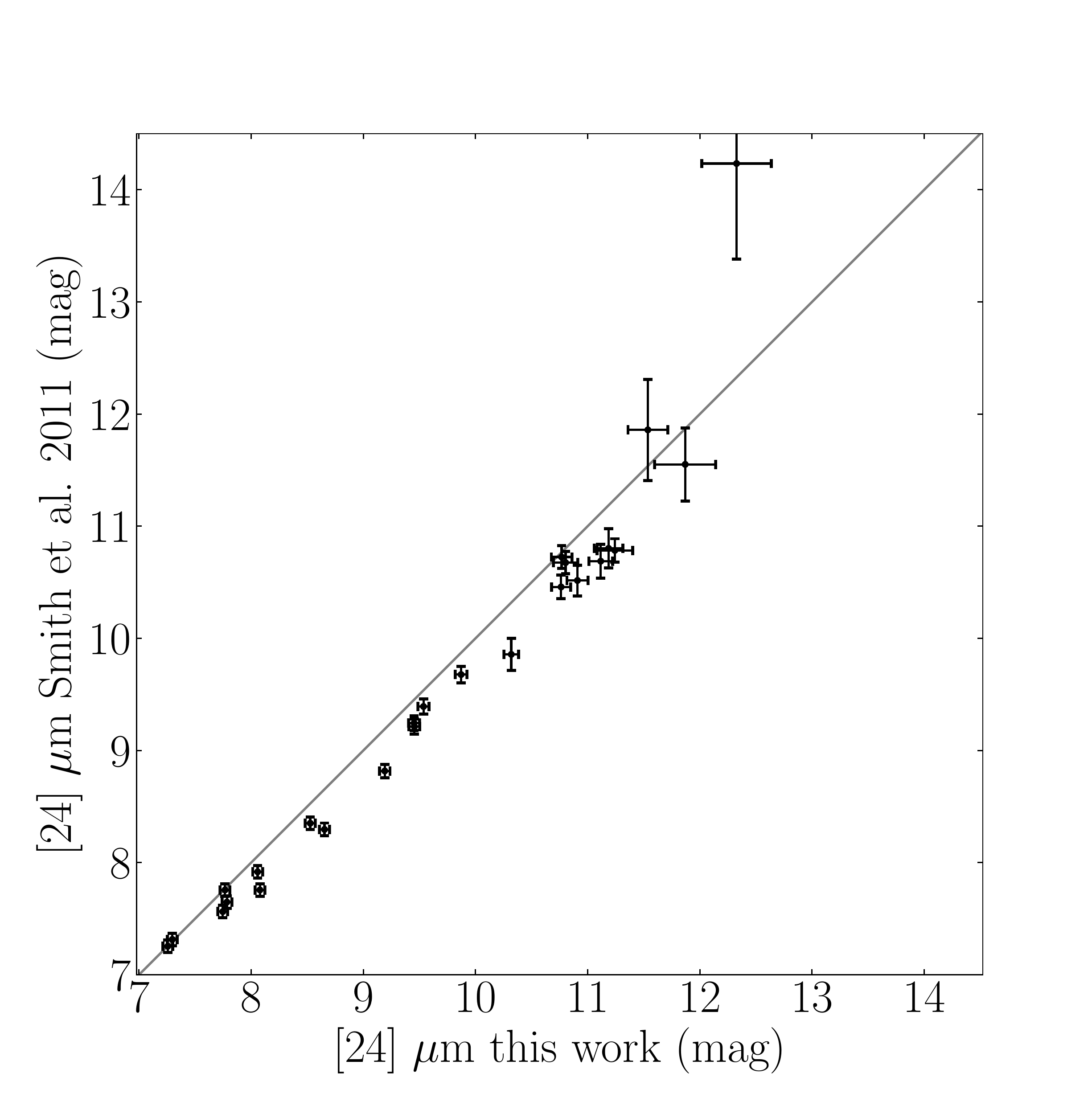}
    \includegraphics[width = \columnwidth]{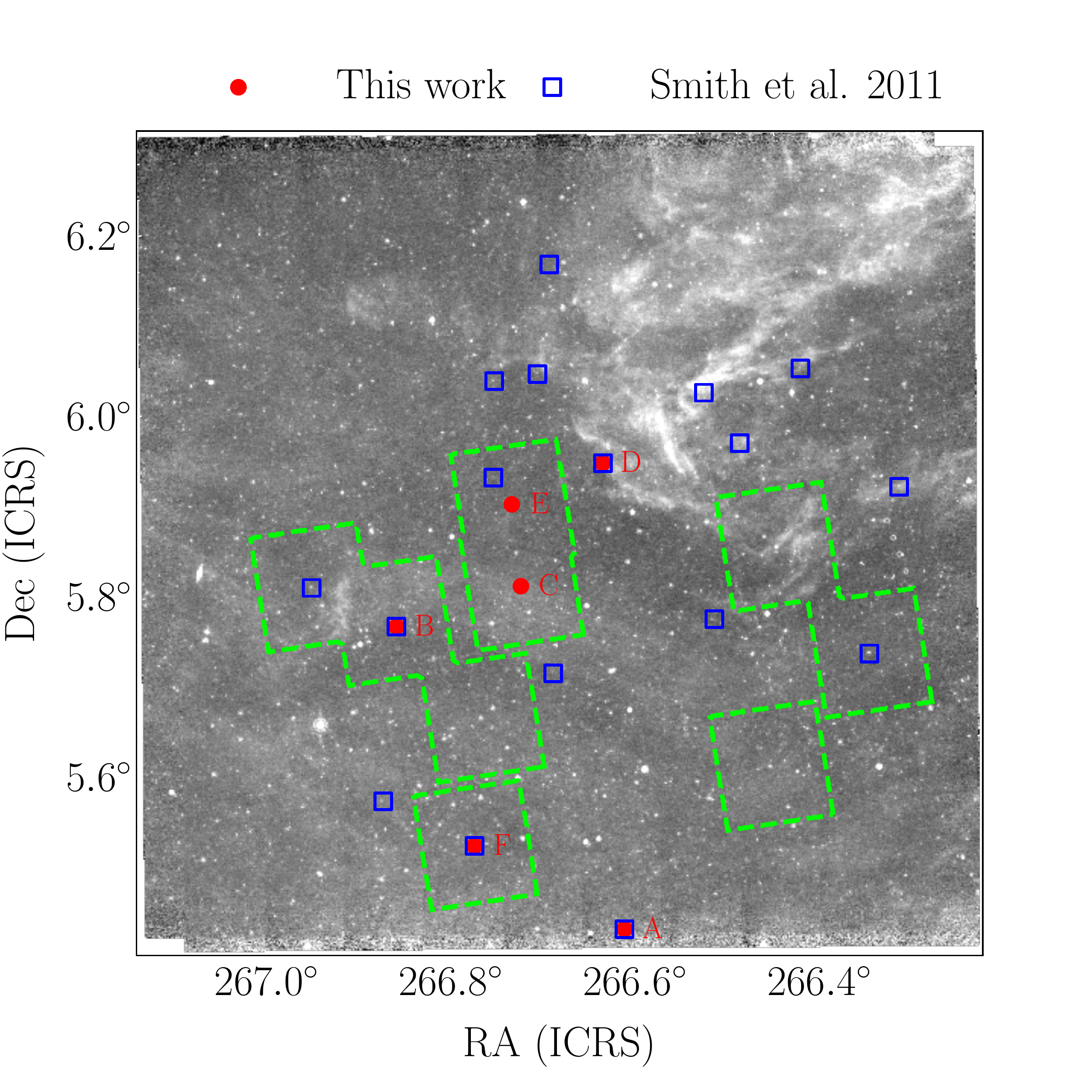}
    \caption{Top: Comparison between the MIPS1 photometry obtained in this work and that published on \citet{Smith+11} for the common sources. Bottom: Spatial distribution of the sources with excess in this work (red dots, labelled as follows, A: 2MASS~J17462472+0517213, B: HD~161733, C: HD~161621, D: TYC~428-1938-1, E: TYC~428-980-1, F: HD~161734) and in \citet{Smith+11} (blue squares). The areas limited by green lines indicate the coverage of the \textit{Spitzer} program ID 3347 which was not included in \citet{Smith+11}. }
    \label{fig:MIPS_comparison}
\end{figure}{}

We used all the optical and IR photometry available in the DANCe catalogue\footnote{\url{http://vizier.u-strasbg.fr/viz-bin/VizieR?-source=J/A+A/631/A57}}, i.e. $G, G_{\rm BP},  G_{\rm RP}, g, r, i, z, y, J, H, Ks$. We also cross-matched (using a cross-match radius of 1\arcsec) our sample with the AllWISE catalogue \citep{Wright+10} and found 704 sources with a counterpart in the $W1$ (3.4~$\mu$m), $W2$ (4.6~$\mu$m), $W3$ (12.1~$\mu$m), $W4$ (22.2~$\mu$m) bands. 

We queried the \textit{Spitzer} Heritage Archive for all the IRAC1 (4.6~$\mu$m), IRAC2 (4.5~$\mu$m), IRAC3 (5.8~$\mu$m), IRAC4 (8.0~$\mu$m) and MIPS1 (24~$\mu$m) data within a radius of 3\degr\, the estimated size of the cluster, around the centre. The program IDs of the observations in this area are given in Table~\ref{tab:spitzer_programID} and the footprints of the various bands are displayed in Figure~\ref{fig:Spitzer_coverage}. The majority of the data come from program ID 40601 which has been analysed in \citet{Smith+11}, but a significant number of images were added from program ID 3347 over part of the area. Our reduction began from the S18.25.0 pipeline-processed artifact Corrected Basic Calibrated Data (CBCD) in the case of IRAC, and from the S18.12.0 pipeline-processed Basic Calibrated Data (BCDs) in the case of MIPS. The self-calibration recommended in the \textit{Spitzer} Data Analysis Cookbook was applied in the case of MIPS to remove artefacts as well as  bright and dark latents present in the BCD images. We then combined these into deep mosaics using the recommended  version 18 of MOPEX (MOsaicker and Point source EXtractor) provided by the \textit{Spitzer} Science Center using the standard parameters (see the MOPEX User’s Guide for details on the data reduction). 

In the case of IRAC, point sources were detected using SExtractor \citep{Bertin+96}, and their PRF-fitting photometry was measured using APEX, the photometry package that is part of MOPEX. According to the manual, the colour corrections tabulated in the IRAC and MIPS handbooks are marginal for our sources ($T_{\rm eff}>4\,000$~K) and we therefore neglected them. 

In the case of MIPS, an extra step was performed before extracting the sources and measuring their photometry. The presence of a bright extended nebulosity (see Fig.~\ref{fig:nebulosity}) indeed compromises the detection and measurements as the background estimations implemented in SExtractor and APEX are not optimised to deal with such extended emission. We therefore applied the nebulosity filter described in \citet{Irwin+10} to the pipeline produced mosaic. A spatially variable Point Spread Function (PSF) was then computed using PSFEx \citep{2013ascl.soft01001B} and the final PSF photometry was extracted using SExtractor again. We verified that the SExtractor PSF and APEX fluxes were in good agreement within the uncertainties, but kept SExtractor measurements as it detected and deblended more sources than APEX. To calibrate the fluxes of SExtractor we used the APEX photometry as reference and computed a linear fit which resulted in a zero point of $140.89\pm0.12$~Jy. We applied this zero point to all the SExtractor fluxes, and its uncertainty was added quadratically to the flux error. To convert fluxes into magnitudes we used the magnitude zero points provided in the instruments handbooks. For IRAC they are 280.9~Jy, 179.7~Jy, 115.0~Jy, and 64.9~Jy for bands 1, 2, 3, and 4, respectively and for MIPS1 it is 7.17~Jy. As explained in the instruments handbooks, the estimated level of accuracy of the photometric measurements is of 3\% for IRAC and 4\% for MIPS1. Therefore, we also added these uncertainties in quadrature to the statistical flux uncertainties. We estimated the global offset of our sources with respect to \textit{Gaia} DR2 and AllWISE to be $\lesssim 2\arcsec$ in all bands inspecting the distribution of cross-match separations. The sources were then cross-identified with our input catalogue using a 2\arcsec\, radius as maximum separation. The number of matches in each case is reported in Table~\ref{table:copleteness}. In Table~\ref{tab:photometry} we provide the photometry used for the sources candidates of hosting a debris disc. An extended version of this table including all the photometric bands for all the members of IC~4665 is available at CDS.

We compared the photometry obtained with the one published in \citet{Smith+11}. The photometric measurements obtained in the four IRAC channels are consistent within the uncertainties except for a few objects close or above saturation. When we compared the MIPS1 photometry we found that for several sources their magnitudes are systematically lower than our measurements (see Fig.~\ref{fig:MIPS_comparison}, top). We checked that our SExtrator PSF fits are good, as demonstrated by the very low levels of residuals and good reduced $\chi^2$. We find several reasons to explain these differences. First, we used a more recent version of the pipeline (both MOPEX and APEX). In particular, it includes a significant background improvement thanks to the self-calibration of the data mentioned above. Second, we used a superior background subtraction with the nebulosity filter compared to the standard pyramidal median filtering used in SExtractor and MOPEX that produces local over-estimations that can severely affect the final photometry in regions of variable extended emission. Indeed, in Figure~\ref{fig:MIPS_comparison} (bottom) we see that several of \cite{Smith+11} sources with 24~$\mu$m excess are in regions of nebulosity. 
We also note that other sources are in regions where our images are more sensitive since we combined the data of two programs (areas limited by green lines). Finally, we emphasise that we provide PSF photometry for all the objects while \citet{Smith+11} measured aperture photometry when their PSF fit failed.

\subsection{Photometry filtering}
\label{subsec:phot_filter}

\begin{table}[ht]
\centering
\caption{Number and percentage of members of IC~4665 detected for each photometric band before and after filtering the photometry. The total number of members is 819. }
\label{table:copleteness}
\begin{tabular}{c|rr|rr}
  \hline
  \hline  
 & \multicolumn{2}{c}{Initial} & \multicolumn{2}{|c}{After filtering}  \\[0.5ex] 
  \hline  
Filter & Num.  & Pct. & Num. &   Pct. \\[0.5ex] 
 \hline
 \hline 
  $G$              &   727  &  89\%   &  727  &  89\%   \\
  $G_\textup{BP}$  &   698  &  85\%   &  698  &  85\%   \\
  $G_\textup{RP}$  &   699  &  85\%   &  699  &  85\%   \\
  \hline
  $g$              &   566  &  69\%   &  566  &  69\%   \\
  $r$              &   695  &  85\%   &  695  &  85\%   \\
  $i$              &   766  &  94\%   &  766  &  94\%   \\
  $z$              &   755  &  92\%   &  755  &  92\%   \\
  $y$              &   789  &  96\%   &  789  &  96\%   \\
  \hline
  $J$              &   815  & 100\%   &  815  & 100\%   \\
  $H$              &   781  &  95\%   &  781  &  95\%   \\
  $Ks$             &   778  &  95\%   &  778  &  95\%   \\
  \hline
  W1               &   704  &  86\%   &  577  &  70\%   \\
  W2               &   704  &  86\%   &  560  &  68\%   \\
  W3               &   704  &  86\%   &  148  &  18\%   \\
  W4               &   704  &  86\%   &   18  &   2\%   \\
  \hline  
  IRAC1            &   218  &  27\%   &  218  &  27\%   \\
  IRAC2            &   219  &  27\%   &  219  &  27\%   \\
  IRAC3            &   204  &  25\%   &  202  &  25\%   \\
  IRAC4            &   200  &  24\%   &  199  &  24\%   \\
  \hline
  MIPS1            &    45  &   5\%   &   45  &   5\%   \\
   \hline
   \hline
\end{tabular}
\end{table}

\begin{figure}
    \centering
    \includegraphics[width = \columnwidth]{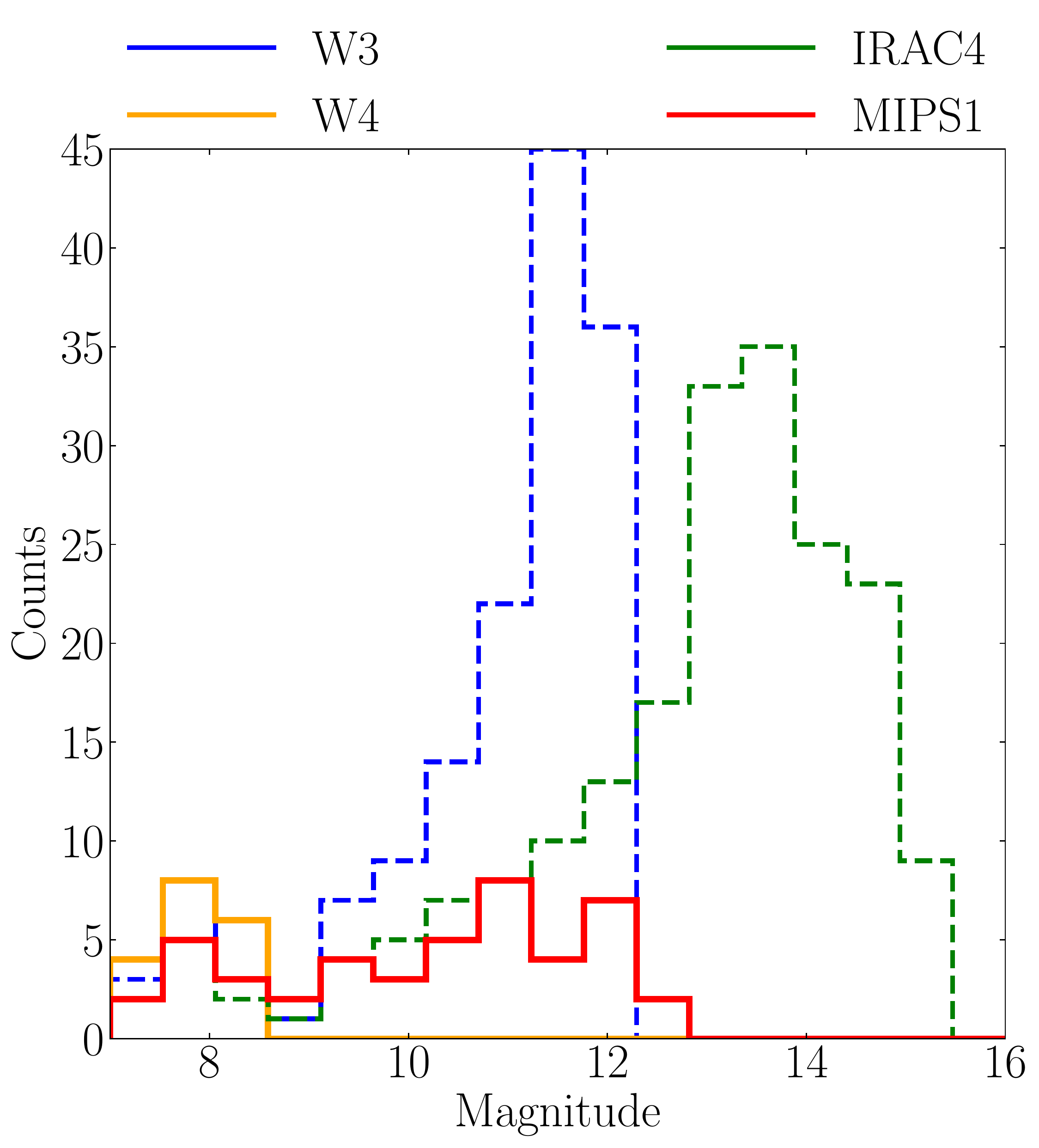}
    \caption{Magnitude distribution of the members of IC~4665 for the filters $W3$ (12~$\mu$m), $W4$ (22~$\mu$m), IRAC4 (8~$\mu$m), and MIPS1 (24~$\mu$m).}
    \label{fig:completeness}
\end{figure}{}

Photometric measurements can be affected by a number of problems (e.g. saturation, blending with a nearby source, cosmic rays, etc.) which can alter the true values and lead to unreliable measurements. Such contaminated photometric measurements can lead to a false IR excess detection or prevent the detection of a real excess. To minimise the impact of dubious photometric measurements, we applied filtering criteria specifically designed for each instrument. 

The WISE photometry is well-known to be affected by a large number of artefacts which are identified and flagged in the AllWISE catalogue. Filtering these sources is essential to discard unreliable photometry. In this work we applied the following filtering:
\begin{itemize}
    \item "cc\_flags". We only keep sources with 0 flag, which means that they are unaffected by any of the known artefacts.
    \item "ext\_flg". We only keep sources with 0 flag, which means that they are point-source objects, excluding extended objects.
    \item "ph\_qual" is the photometric quality flag. We consider flags A (SNR $\geq10$) and B ($3<$ SNR $<10$) as good quality photometry, flags C ($2<$ SNR $<3$) as bad quality photometry, and flag U (SNR $<2$) as an upper limit. The bad quality photometry and the upper limits are shown throughout our analysis but are not considered as reliable measurements. 
\end{itemize}{}

For \textit{Spitzer} observations, we discarded any detection with SNR~$<3$. Detections with $3<$~SNR~$<5$ are considered to be marginal and they should be considered with caution. 

In Table~\ref{table:copleteness} we report the number of sources detected in each band before and after the filtering process. In spite of their better sensitivity, the \textit{Spitzer} images include less sources simply because they cover a smaller area (1$\times$1\degr, see Fig.~\ref{fig:Spitzer_coverage}). Another important remark is that WISE detections are the most affected by our filtering criteria, specially at large wavelengths ($W3$ and $W4$) where many sources are only detected as upper limits. In Figure~\ref{fig:completeness} we see the magnitude distribution of the members of IC~4665 in the mid-infrared photometric bands. There are 45 sources with MIPS1 photometry in the magnitude range 7.2--12.4~mag. The detection limit of the $W4$ channel at 22~$\mu$m is slightly shallower and there are 18 sources in the magnitude range between 7--8.6~mag.

\subsection{Completeness}\label{subsec:completeness}

One of the goals of this study is to measure the debris disc fraction of IC~4665. For that, defining a sample for which the photometric surveys are complete is crucial to interpret the results and compare with other studies. Establishing the completeness of photometric surveys can be a complex task, especially when variable extended emission is present (see Fig.~\ref{fig:nebulosity}).

In the following, we derive an estimate of the completeness limits of the photometric filters MIPS1 and $W4$. We take the maximum of the magnitude distribution of all the sources in the field of view (members and field stars) as the completeness limit. These distributions show a maximum around 11.8~mag for MIPS1 and around 8.5~mag for $W4$. We find no significant variations in the areas affected by the extended emission and, as a first approximation, use these numbers for the entire survey. To convert the completeness in magnitude to a fundamental parameter such as the mass, we used the BT-Settl atmospheric models of \citet{Allard14}. At the age and distance of IC~4665, we find that the magnitude limit of MIPS1 corresponds to a temperature of 4\,000~K and a mass of 0.75~M$_\sun$. This corresponds to a spectral type of mid-K (Table~A5 from \citealt{Kenyon+95}). The BT-Settl models do not cover the hottest objects and can not be used to convert the $W4$ completeness limit in magnitude into a mass. Therefore, we followed a different approach and transformed the $W4$ magnitude to a $G-Ks$ colour using an empirical relation defined by the members of IC~4665. The limiting magnitude of $W4=8.5$~mag corresponds to an intrinsic $G-Ks~\sim0.2$~mag (assuming an extinction A$_\textup{V}=0.46$~mag). This color corresponds to an effective temperature of 8\,400~K and a mass of 1.75~M$_\sun$ according to the PARSEC-COLIBRI models \citep{Marigo+17}. This is equivalent to a spectral type of mid-A \citep{Kenyon+95}.

%

\section{Infrared excess detection}
\label{sec:colour-colour-diagrams}

In order to identify cluster members with debris discs, we use colour-colour diagrams to detect IR excesses at different wavelengths, a method that is entirely empirical. We chose to analyse the WISE and \textit{Spitzer} data independently because of their significantly different wavelength coverage, spatial resolution and sensitivities.

\subsection{MIPS 24~$\mu$m data}
\label{subsec:MIPS}

\begin{figure}
    \centering
    \includegraphics[width = \columnwidth]{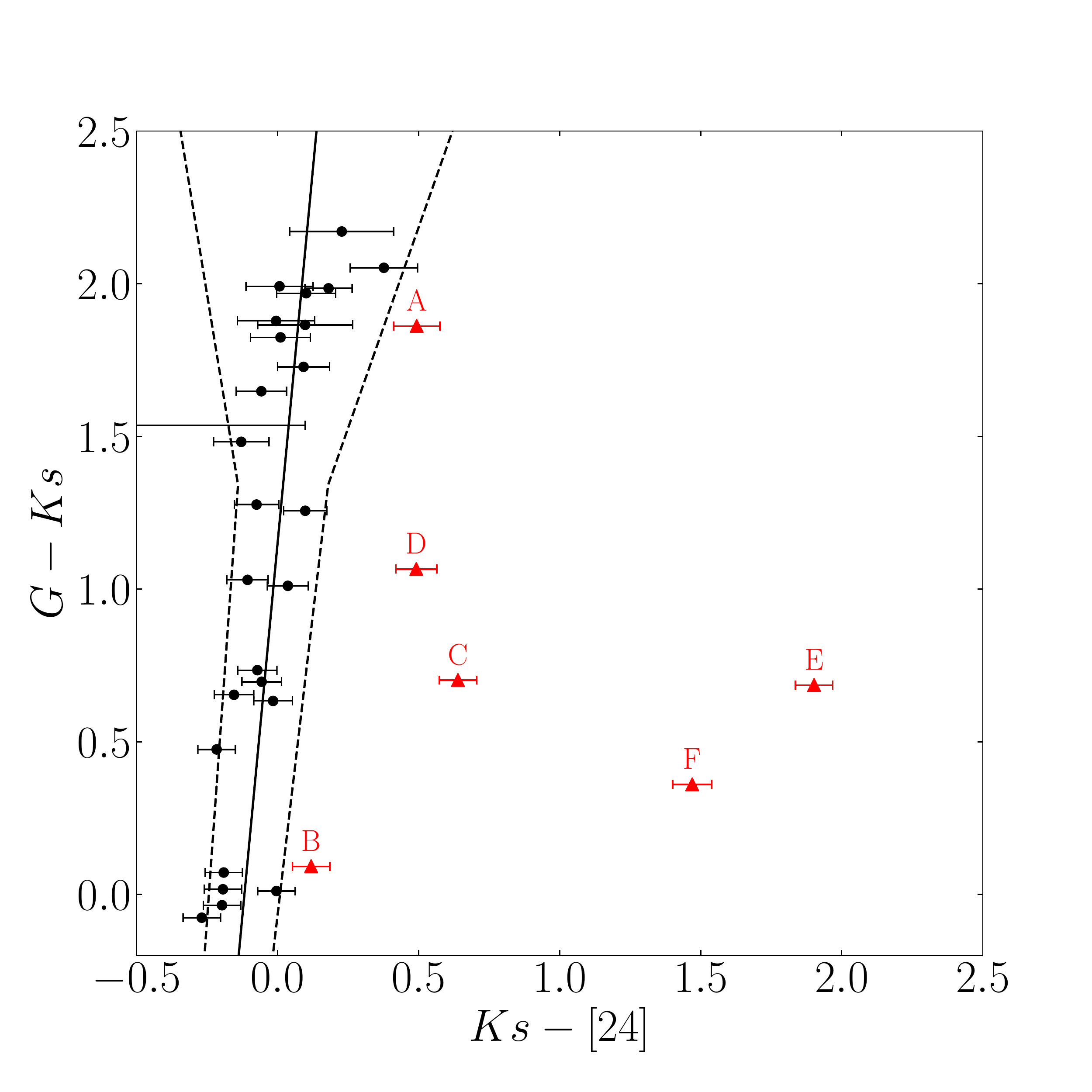}
    \caption{$(G-Ks)-(Ks-[24])$ colour-colour diagram of the members of IC~4665 with 24~$\mu$m photometry. The locus of photospheric emission is indicated by the solid line and the $3\sigma$ uncertainties are represented by the dashed lines. The members with IR excess are indicated as red triangles, the labelling code is the same as caption of Fig.~\ref{fig:MIPS_comparison}}.      \label{fig:24_excess}
\end{figure}{}

\begin{figure*}
    \centering
    \includegraphics[width = \textwidth]{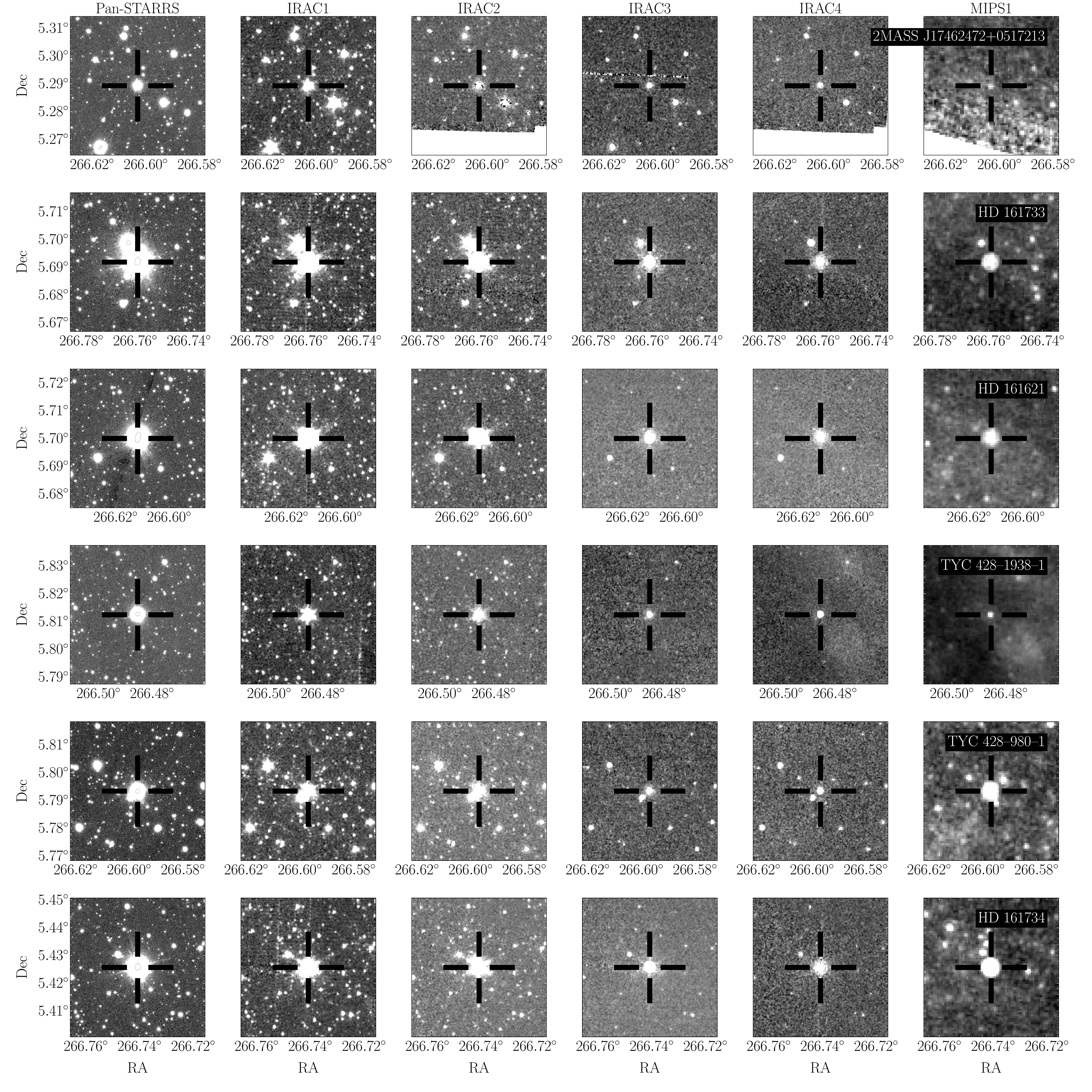}
    \caption{Multifilter \textit{Spitzer} images of the sources with IR excess.}
    \label{fig:images-Spitzer}
\end{figure*}{}

The photometric colour $Ks-[24]$ has been commonly used in the literature to detect sources with 24~$\mu$m excess emission \citep[e.g.][]{Gorlova+06, Stauffer+10}. We followed the methodology of these works and used colour-colour diagrams to discern between the excess and non-excess population. The only difference is that we used the photometry of the recent \textit{Gaia} DR2 $G$ filter which is more precise, uniform, and extended than the Johnson $V$ filter. 

In order to detect sources with IR excess, it is essential to first outline the location of the sources with photospheric colours (i.e. those which do not have an IR excess). We used the \textit{Spitzer} observations in the Pleiades of \citet{Gorlova+06} to define the photospheric sequence in the $(G-Ks)-(Ks-[24])$ colour-colour diagram. We note that this is the same approach that \citet{Gorlova+07} used for the 30--40~Myr cluster NGC~2547. Indeed, several authors have found similar relations in young clusters (e.g. \citealt{Stauffer+10} for the Hyades, and \citealt{Plavchan+09} for nearby young stars). We used only the reliable 24~$\mu$m photometry for the Pleiades sources classified as not having excess by the authors\footnote{sources in their Table~2 with no asterisk.}. Equivalently to what they did, we fitted a linear polynomial relation (see Fig.~\ref{fig:gorlova+06}) and obtained:
\begin{equation*}
    (Ks-[24]) = 0.102(\pm0.013)\times(G-Ks) - 0.12 (\pm0.02)
\end{equation*}{}

We used this relation to define the photospheric emission locus in our data. To establish a confidence interval in which we believe that there is no emission excess, we fitted a segmented linear function with two slopes to account for the fact that at high SNR the measures are dominated by photon noise and at low SNR by Poisson noise (see Fig.~\ref{fig:fit_eMIPS1}). The point of change in trend is a free parameter of the fitting process.

\begin{equation*}
    \textup{e}[24] = \begin{cases} c1 + k1\times(G-Ks), \qquad (G-Ks)<1.34\\ 
    c2 + k2\times(G-Ks), \qquad (G-Ks)\geq 1.34\end{cases}
\end{equation*}{}
The parameters of the fit are $c1 = 0.042(\pm0.018)$~mag, $c2 = -0.07(\pm0.05) $~mag, $k1=0.008(\pm0.011)$, and $k2=0.09(\pm0.03)$.

We consider that sources with a colour $Ks-[24]$ larger (within the uncertainties) than the photospheric emission locus plus the $3\sigma$ uncertainties of the 24~$\mu$m photometry have an excess in the 24~$\mu$m emission. According to this criterion, we find 5/45 candidates with excesses in the 24~$\mu$m emission. Three (HD~161733, TYC~428-1938-1, and 2MASS~J17462472+0517213) had already been reported by \citet{Smith+11} and the other two (HD~161621 and TYC~428-980-1) are new candidates. In Figure~\ref{fig:images-Spitzer} we see that all the candidates have a clear detection in all the \textit{Spitzer} photometric bands. 

In the case of  \citet{Smith+11}, they detected 14 additional sources with 24~$\mu$m excesses not present in our analysis. Four of them are simply not classified as members by \citet{Miret-Roig+19} and thus were not considered in this study. We have checked that three out of these four do not show any excess at 24~$\mu$m according to our photometry. The fourth object (HD~161734) has WISE+\textit{Spitzer} excesses in our photometry (see Figs.~\ref{fig:24_excess}, \ref{fig:colour-colour-IRAC}, and \ref{fig:colour-colour-WISE}) but it had a low membership probability (p = 0.002\%) with \textit{Gaia} in \citet{Miret-Roig+19} and therefore was not initially considered in this work. We refer to Sect.~\ref{sec:candidates} for a more detailed discussion of this object. For the remaining 10 sources the MIPS 24\,$\mu$m photometry is significantly discrepant between the two studies (as shown in Fig.~\ref{fig:MIPS_comparison}), which explains why they found an excess and we did not. 

\subsection{IRAC 3.6--8.0~$\mu$m data}
\label{subsec:IRAC}

\begin{figure*}
    \centering
    \includegraphics[width = \textwidth]{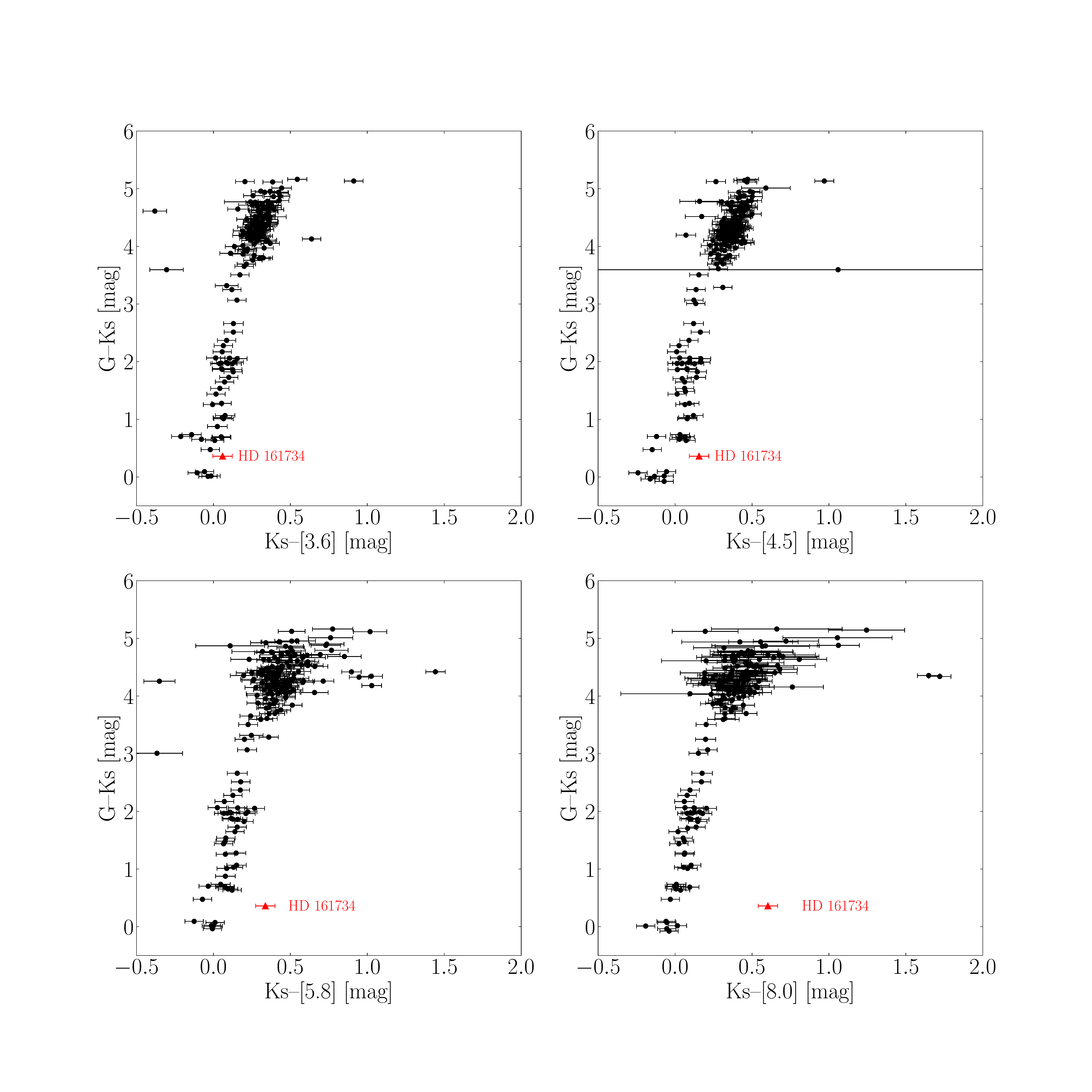}
    \caption{Colour-colour diagrams of the IRAC channels. Black dots indicate sources with high signal-to-noise ratio (SNR$>$5) and the red triangle indicates the source with IR excess.}
    \label{fig:colour-colour-IRAC}
\end{figure*}{}

Similarly to what we did in Sect.~\ref{subsec:MIPS} with the MIPS photometry, now we use the four channels of IRAC to search for possible candidates with a near-infrared excess. In Figure~\ref{fig:colour-colour-IRAC} we represent the colour-colour diagrams $Ks-[3.6], Ks-[4.5], Ks-[5.8], Ks-[8.0]$ against $G-Ks$. None of our members have an excess on any of the IRAC bands. There are a few points in each colour-colour diagram which seem to be much redder than the mean photospheric locus but none of them show an excess in two consecutive bands. These are spurious or blended detections in the images, or have a high $\chi^2$ of the PSF fit. 

The source HD~161734 shows an increasing excess in the colour-colour diagrams of the IRAC channels. This excess was also detected by \citet{Smith+11} in the 5.8 and 8.0~$\mu$m channels. 

\subsection{WISE 3.4--22.2~$\mu$m data}
\begin{figure*}
    \centering
    \includegraphics[width = \textwidth]{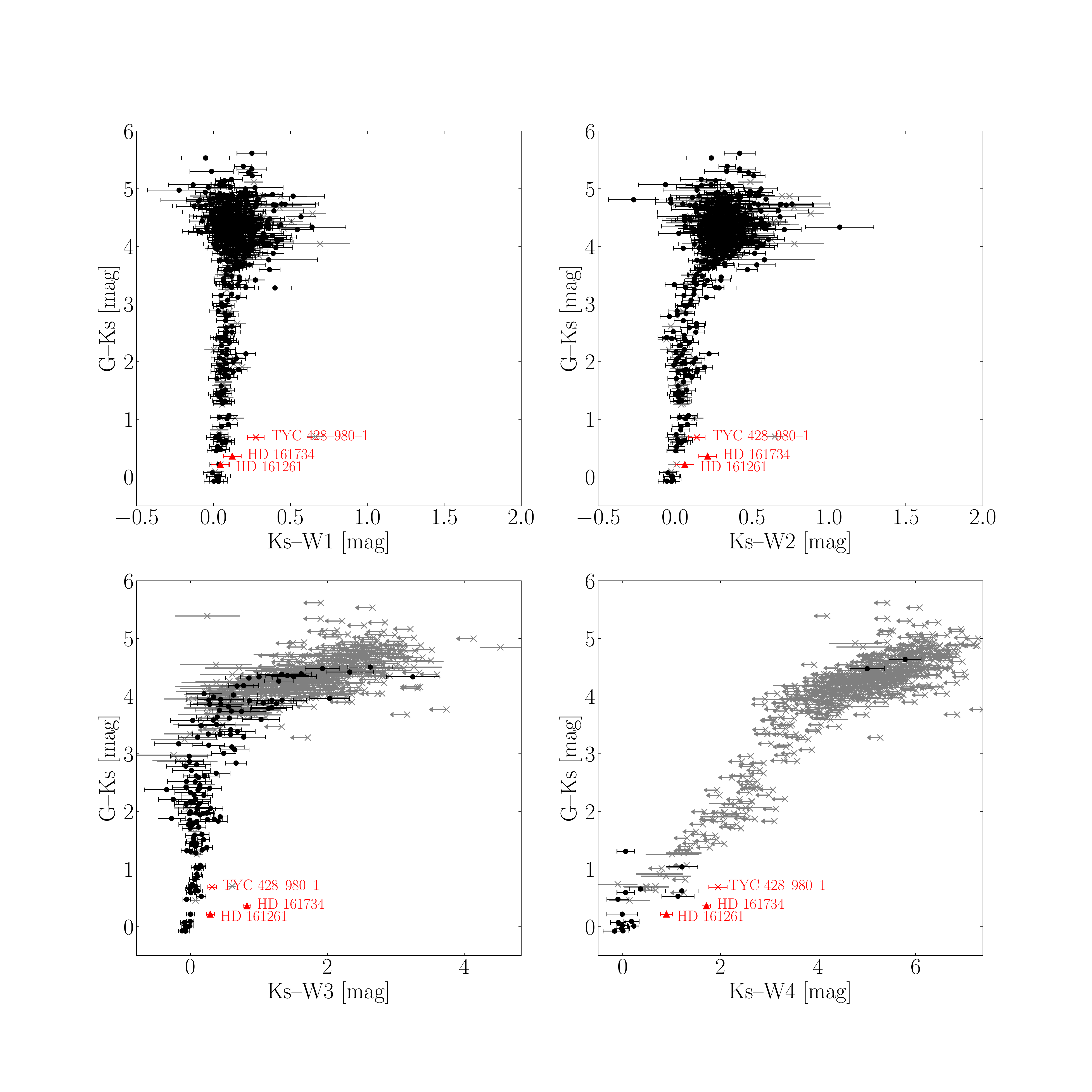}
    \caption{Colour-colour diagrams of the WISE photometric bands. Black dots indicate sources with good photometric quality, grey crosses indicate low quality photometry or upper limits, and red triangles indicate an IR excess. The source TYC~428--980--1 (red cross) shows an IR excess but is flagged as an extended object (see text).}
    \label{fig:colour-colour-WISE}
\end{figure*}{}

\begin{figure*}
    \centering
    \includegraphics[width =0.95 \textwidth]{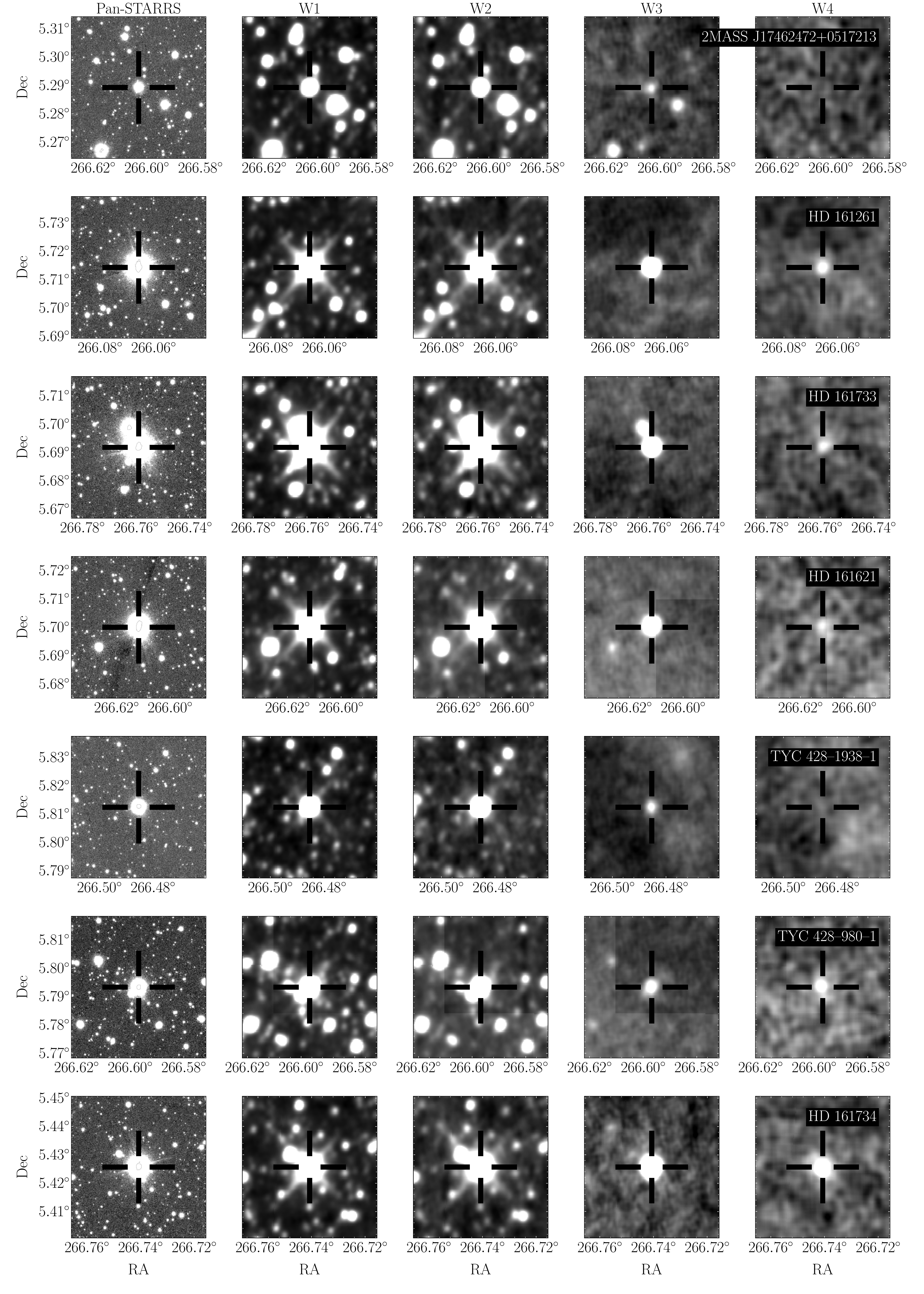}
    \caption{Multifilter WISE images of the IC~4665 sources with IR excesses.}
    \label{fig:image-WISE}
\end{figure*}{}

Similarly to what we did in Sect.~\ref{subsec:MIPS} for MIPS and in Sect.~\ref{subsec:IRAC} for IRAC, here we present colour-colour diagrams using the WISE photometry (see Fig.~\ref{fig:colour-colour-WISE}). The photometric measurements classified as "good" are represented by black dots, according to the criteria defined in Sect.~\ref{subsec:phot_filter}. 
In these panels we have also represented data classified as "bad" to illustrate the loss of sensitivity at longer wavelengths, resulting in the overlap of good and bad quality data in $W3$ (mainly for the coldest objects) and $W4$ (almost along the whole cluster sequence). In the bands $W1$ (3.4~$\mu$m) and $W2$ (4.6~$\mu$m) we do not see any source with a significant excess, with the exceptions of TYC~428-980-1 which is flagged as a extended source by the WISE catalogue (probably due to the blending of nearby sources, see Fig.~\ref{fig:image-WISE}), and HD~161734 which shows an increasing excess in all the WISE bands.

In addition to HD~161734, in the $W3$ (12~$\mu$m) band we see two objects redder than the photospheric sequence defined by the majority of sources. One is HD~161261, a good candidate to host a debris disc since it also displays an excess in $W4$ (22~$\mu$m). Moreover, the WISE images of this source look clean (see Fig.~\ref{fig:image-WISE}). We note that this source is not covered by \textit{Spitzer} because is not in the central $1\times 1\degr$ area. The second source is TYC~428-980-1 which, as mentioned above, was initially discarded for being flagged as an extended object by the WISE catalogue. However, if we look at the WISE images (Fig.~\ref{fig:image-WISE}), we see that the source appears blended in the $W1$ and $W2$ bands, but not in $W3$ and $W4$, where the object shows an increasing excess. In Section~\ref{sec:candidates} we discuss this object more in detail, and explain if the nearby source detected at shorter wavelengths might be unresolved in $W3$ and $W4$, with the subsequent contamination of the derived photometry.

As commented above, in bands $W3$ and $W4$ there is a significant increasing amount of "bad" photometric measurements and upper limits. This is due to the limited sensitivity of WISE, and the increasing amount of diffuse emission present at those wavelengths. 

There are a few sources at the limit of sensitivity which have not been flagged by WISE as "bad" and seem to have an excess. We have checked the images and discarded them because they do not show any detected source.

\section{SEDs}
\label{sec:SED}

\begin{figure*}
    \centering
    \includegraphics[width = \textwidth]{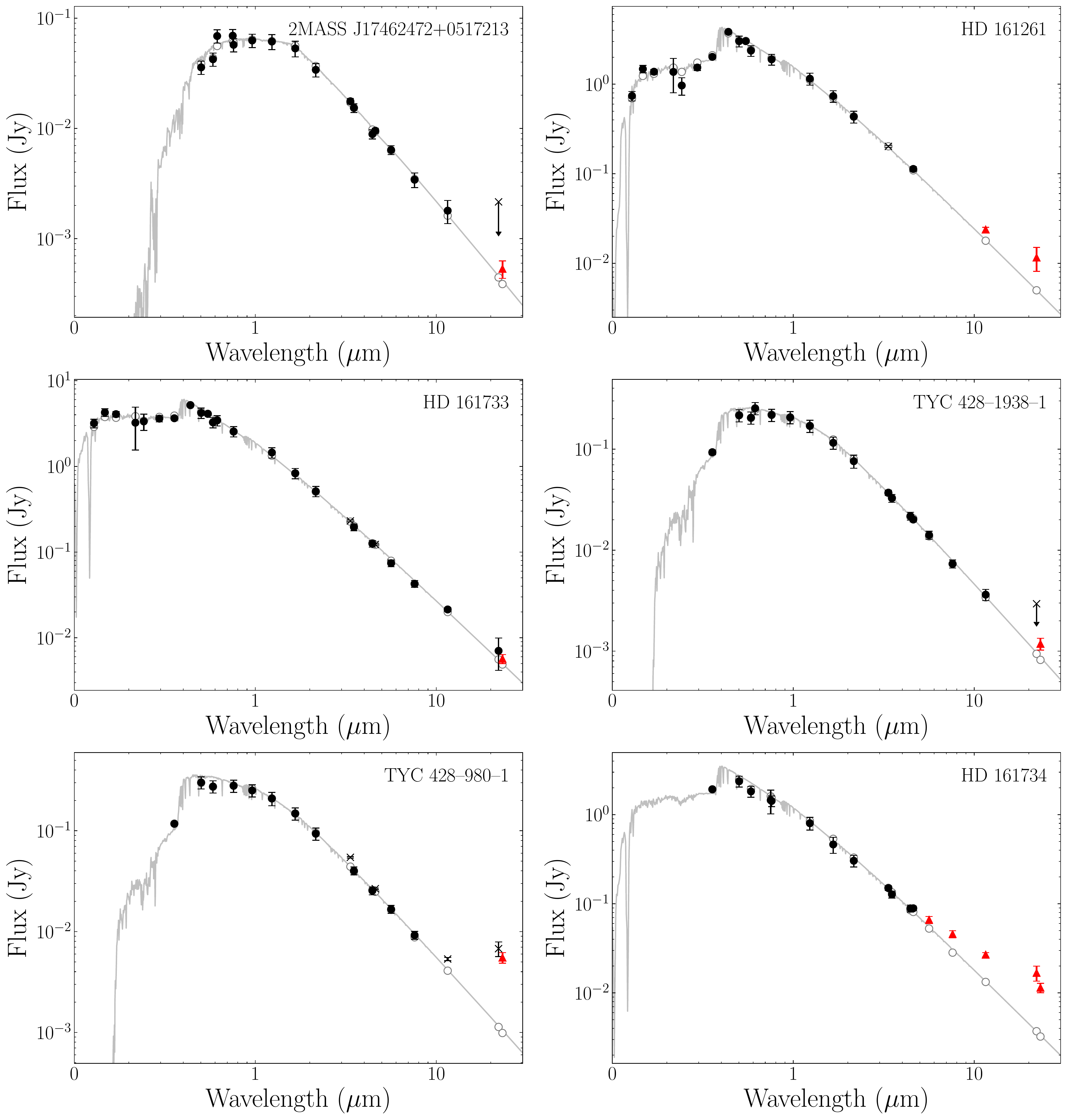}
    \caption{SEDs of the sources for which we have detected an IR excess. The black dots indicate the photometric measurements used in the fit, the grey crosses indicate the photometric measures not used in the fit (either low quality photometry or upper limits), the red triangles indicate a photometric excess detected by VOSA. We report the 3$\sigma$ uncertainties, some of which are smaller than the markers. We note that the DANCe photometry has a minimum uncertainty of 0.05~mag. The grey line represents the photospheric best fitted Kurucz model.  }
    \label{fig:SED}
\end{figure*}{}

\begin{landscape}
\setlength{\tabcolsep}{4pt} 
\renewcommand{\arraystretch}{1.5} 
\begin{table}
\centering
\caption{List of candidates for which we have detected an IR excess. Columns indicate: (1) Object ID; (2) spectral type; (3) Bayesian distance; (4--5) \textit{Gaia} DR2 $G$-band magnitude and extinction;  (6--7) photometric effective temperature from \textit{Gaia} DR2 and \citet{Smith+11}; (8--12) extinction, effective temperature and the 96\% confidence level, $\log g$, and reduced $\chi^2$ parameter of the best fit SED; (13--14) significance of the $W4$ and 24~$\mu$m excess defined as $\left(\frac{F_{obs}-F_{phot}}{\sigma_{obs}}\right)^2$; (15) binary flag (from literature); (16) new candidate to host a debris disc. ${^*}$We note that eventually we reject the source HD~161621 as a candidate (see text). }
\label{tab:candidates}
\begin{tabular}{|l| r r c c c c| c r r r r | r r | c c|}
\hline
\hline
  \multicolumn{1}{|c|}{Object} &
  \multicolumn{1}{c}{SpT (Ref.)} &
  \multicolumn{1}{c}{$d$} &
  \multicolumn{1}{c}{$G$} &
  \multicolumn{1}{c}{A$_{G}$} &
  \multicolumn{1}{c}{T$_\textup{eff, GDR2}$} &
  \multicolumn{1}{c}{T$_\textup{eff, S11}$} &
  \multicolumn{1}{|c}{A$_{V,~\textup{SED}}$} &
  \multicolumn{1}{c}{T$_\textup{eff, SED}$} &
  \multicolumn{1}{c}{T$_\textup{eff, SED} ~ 96\%~\textup{CL}$} &
  \multicolumn{1}{c}{$\log g$} &
  \multicolumn{1}{c|}{$\chi^2$} &
  \multicolumn{1}{c}{$S_{W4}$} &
  \multicolumn{1}{c}{$S_{24}$} &
  \multicolumn{1}{|c}{Binary}  &
  \multicolumn{1}{c|}{New} \\
  
  \multicolumn{1}{|c|}{} &
  \multicolumn{1}{c}{} &
  \multicolumn{1}{c}{(pc)} &
  \multicolumn{1}{c}{(mag)} &
  \multicolumn{1}{c}{(mag)} &
  \multicolumn{1}{c}{(K)} &
  \multicolumn{1}{c}{(K)} &
  \multicolumn{1}{|c}{(mag)} &
  \multicolumn{1}{c}{(K)} &
  \multicolumn{1}{c}{(K)} &
  \multicolumn{1}{c}{} &
  \multicolumn{1}{c|}{} &
  \multicolumn{1}{c}{} &
  \multicolumn{1}{c}{} &
  \multicolumn{1}{|c}{} &
  \multicolumn{1}{c|}{} \\
\hline
              \object{HD 161261} & B9 (1)        & $356\pm17$ &  8.26 & 0.62 & 9\,132 & 8\,842 & 0.6  & 12\,000 & 11\,250--13\,000 & 4.5 & 10.0 &  32.8 &       & ? & Y \\
              \object{HD 161733} & B7(2)--B8 (1) & $322\pm13$ &  7.96 & 0.43 & 9\,124 & 8\,740 & 0.65 & 15\,000 & 14\,000--16\,914 & 5.0 &  5.2 &   2.3 &  12.6 & ? & N \\
              \object{HD 161621} & A0 (1)        & $314\pm11$ &  9.45 & 0.61 & 7\,886 &        & 0.4  &  8\,250 &  7\,250--11\,000 & 4.0 &  9.0 &       &       & Y & Y$^{*}$ \\ 
         \object{TYC 428-1938-1} & A  (3)        & $338\pm15$ & 11.01 & 0.77 & 6\,197 & 7\,840 & 0.7  &  7\,500 &   6\,031--8\,463 & 4.5 &  0.9 &       &  47.4 & N & N \\
          \object{TYC 428-980-1} & A2 (3)        & $340\pm11$ & 10.36 &      & 7\,310 &        & 0.35 &  7\,750 &  6\,750--10\,250 & 4.0 &  0.6 &  25.0 & 417.1 & N & Y \\
\object{2MASS J17462472+0517213} & G2--G3 (4)    & $336\pm10$ & 12.67 & 0.78 & 5\,155 & 5\,459 & 0.65 &  5\,750 &   5\,000--6\,500 & 5.0 &  3.8 &       &  20.8 & N & N \\
     \object{HD 161734}          & B8 (5)--A0 (1)& $447\pm20$ &  8.82 & 0.75 & 8\,478 & 8\,420 & 0.9  & 13\,000 & 10\,705--14\,000 & 4.5 &  5.9 & 150.0 & 320.3 & ? & N \\
\hline
\hline\end{tabular}
\tablebib{ (1)~\citet{Cannon+93}; (2)~\citet{Kraicheva+80}; (3)~\citet{Giampapa+98}; (4)~\citet{Allain+1996}; (5)~\citet{Gray+02}}

\end{table}
\end{landscape}

We used the Spanish Virtual Observatory (SVO) tool VOSA\footnote{\url{http://svo2.cab.inta-csic.es/theory/vosa/}} \citep[VO SED analyser, ][]{Bayo+08} to further investigate the sources that show significant excesses in the colour-colour diagrams. We use VOSA to fit their SED in order to (i) confirm if the mid-IR excesses are automatically detected by the tool, and (ii) characterise the central sources more in detail.

We used all the photometry described in Sect.~\ref{sec:data} excluding the filtered photometry, upper limits, and saturated measures. In addition, we used the VOSA interface to search for all the Ultraviolet (UV) photometry available. Since several targets are early-type stars, the filters in the bluest part of the spectrum are important to complete the SED. We computed Bayesian distances inferred using \textit{Kalkayotl}\footnote{\url{https://github.com/olivares-j/kalkayotl}} and the \textit{Gaia} DR2 parallaxes. The resultant values are reported in Table~\ref{tab:candidates}. The extinction towards IC~4665 is not yet well constrained. The 3D dust map from \citet{Green+19} reports an extinction of $A_\textup{V}=0.46^{+0.12}_{-0.06}$~mag at the position and distance of the cluster. Recently, \citet{Miret-Roig+19} estimated a median extinction of $A_\textup{V}=0.72$~mag with the \textit{Gaia} DR2 \texttt{a\_g\_val} and \citet{Anders+2019} determined individual extinction values for our targets between 0.09--0.74~mag. In consequence, we let the extinction as a free parameter between 0--1~mag. 

To find the best model which reproduces the observed SEDs, we used the option "Chi-square Fit". This option calculates the synthetic photometry from theoretical spectra for the filters with observed data, and applies a statistical test to find the model that best reproduces the data. The fitting algorithm is able to detect possible IR excesses and then, these points are no longer considered in the final fit.
We used the theoretical atmospheres models of Kurucz (ODFNEW/NOVER models, \citealt{Castelli+97}) with solar metallicity. We allow VOSA to find the best $\log g$ in the interval 4--5, and the best temperature in the range $5\,000<T_\textup{eff}<20\,000$~K.

In Figure~\ref{fig:SED}, we present the observed SEDs together with the best fit models, and in Table~\ref{tab:candidates} we summarise some of their parameters. We see that VOSA independently finds an IR excess for six of the seven candidates for which an excess was detected in colour-colour diagrams (Sect.~\ref{sec:colour-colour-diagrams}). The excess of HD~161733 is at the limit of the VOSA detection ($\gtrsim3\sigma$, see Sect.~\ref{sec:candidates}).
In Table~\ref{tab:candidates} we report the effective temperature measured with different techniques. The effective temperatures from \textit{Gaia} DR2 (T$_\textup{eff, GDR2}$) are computed from the three photometric bands of \textit{Gaia}, and two colours which can be strongly correlated. In addition, they have been obtained with a machine learning algorithm only trained on the range 3\,000--10\,000~K. Stars outside this range (which is the case of several of our candidates) can be systematically under- or overestimated \citep{Andrae+18, GaiaCol+18}. The effective temperatures from \citet{Smith+11} (T$_\textup{eff, S11}$) are obtained from the $B-V$ intrinsic colours through a relation provided by the authors. 
The differences between these two temperatures are of a few hundreds of Kelvin, similar to what \citet{Andrae+18} found in the validation of the \textit{Gaia} DR2 effective temperatures. Additionally, we provide the effective temperatures of the theoretical atmospheric model which best fits the observed SED (T$_\textup{eff, SED}$). In all cases, the temperatures from the SED fitting have higher values with respect to the two photometric temperatures, specially for the two hottest objects which have differences of thousands of Kelvins. In these cases, we trust more the effective temperatures from our SED fitting because they rely on a larger amount of photometric measurements from different instruments (i.e. not correlated), covering a large fraction of the spectra (UV-IR), and are derived using the individual parallaxes for each object. Additionally, our effective temperatures match better the spectral types determined in the literature (see second column of Table~\ref{tab:candidates}).

\section{Candidates of hosting a debris disc}
\label{sec:candidates}
We detected seven stars with mid-IR excesses in one or several of the \textit{Spitzer} and WISE bands. In this section, we discuss them one by one. 

\subsection{\object{HD 161261}} 
\object{HD 161261} displays excesses in $W3$ and $W4$ and the images show a clear detection in both cases. This source is not in the $1\times1\degr$ central region of the cluster and thus is not covered by {\em Spitzer}. It is a new debris disc candidate. The effective temperature of the best theoretical atmospheric model is $12\,000$~K, significantly higher than the values obtained by \textit{Gaia} DR2 and \citet{Smith+11}. However, it matches well with the spectral class B9 from \citet{Cannon+93}. 

Interestingly, this source is classified as a rotating ellipsoidal variable by the General Catalogue of Variable Stars \citep{Samus+17}, which implies that it could be a close binary system. However, we find neither any confirmation of the existence of this binary system nor any hint on the properties of the possible companion.  

\subsection{\object{HD 161733}} 
\object{HD 161733} shows a small excess in MIPS 24~$\mu$m data (see Fig.~\ref{fig:24_excess}). It is also detected in $W4$, but the uncertainties of the photometry are large hindering a confirmation of this excess with WISE. 
The best fit SED corresponds to a model of T$_\textup{eff}=15\,000$~K, significantly hotter than the photometric temperatures of \textit{Gaia} DR2 and \citet{Smith+11}, but consistent with the strong Helium lines present in its spectrum (e.g. Levato \& Mararoda 1977, Hubrig \& Mathys 1996), and the 
spectral classification found in the literature \citep[e.g.][]{Kraicheva+80, Cannon+93}.
The algorithm of VOSA did not detect an infrared excess for this source when both $W4$ and MIPS1 are considered. We checked that the MIPS 24~$\mu$m observation shows a significance excess of 
$3.5\sigma$, at the limit of the algorithm detection. In fact, if we neglect the $W3$ and $W4$ photometry ($W1$ and $W2$ are filtered in Sect.~\ref{subsec:phot_filter}), VOSA automatically detects an excess with MIPS. \citet{Smith+11} also detected a 24~$\mu$m excess for this source. 

The multiplicity of this source has been discussed in different works and the results are not conclusive.
It was included in the catalogue of spectroscopic binaries of \citet{Kraicheva+80}, based on radial velocity studies \citep{Abt+1972, Pedoussaut+1973}.
However, the works of \citet{Crampton+1976} and \citet{Morrell+1991} did not report any RV variability after the analysis of several spectra. According to the work of \citet{Kraicheva+80}, the masses of the primary and the secondary are 4.3~M$_\sun$ and 0.8~M$_\sun$, respectively. This mass ratio cannot explain the excess we observe since the contribution of a 0.8~M$_\sun$ star on the SED of a 4.3~M$_\sun$ star is negligible.
These authors also determined the orbital period of the system $7.3\pm0.8$~days and a separation of $\sim1.4$~AU. 
However the nature of the spectroscopic binary is not confirmed and more observations are needed to characterise this source.

\subsection{\object{HD 161621}}\label{subsec:387} 

\object{HD 161621} is a visual binary star with a separation of 3.2\arcsec \citep{Mason+01}. The companion is the source \object{TYC~428-1977-1}. 
The two companions have very similar magnitudes ($G=9.45$~mag; $G=9.59$~mag), effective temperatures (7\,886~K; 7\,811~K), and parallaxes ($\varpi=3.17\pm0.05$~mas; $\varpi=3.29\pm0.04$~mas) indicating it is an equal mass binary.

This system has a WISE counterpart at a separation of 1.6\arcsec\ from the primary source, slightly larger than our initial search so we added the WISE photometry manually. This source is flagged as an extended object by WISE and we checked that it is not resolved by this instrument. In fact, this system is resolved by \textit{Gaia} and IRAC, and unresolved by the International Ultraviolet Explorer (IUE), WISE and MIPS. 2MASS detects the two components but the photometry is flagged as contaminated. 

We only used the photometry which resolved the system (\textit{Gaia} and IRAC) to fit a SED (see Fig.~\ref{fig:SED-387}). We can see that all the unresolved channels (crosses) provided a photometric measure systematically brighter than the predicted by the model. Our SED fit shows that the excess we detected with MIPS is in fact due to the companion since the excess is of 0.75~mag, exactly what we expect for an equal mass binary. For this reason, we no longer consider this source as a candidate to host a debris disc.

\subsection{\object{TYC 428-1938-1}} 
\object{TYC 428-1938-1} displays an excess in 24~$\mu$m data which was already reported by \citet{Smith+11}. This source has WISE photometry but the sensitivity of $W4$ is too low to be detected in that channel (see Fig.~\ref{fig:image-WISE}) so we cannot use WISE to confirm this excess. The SED shows a 24~$\mu$m excess with respect to the photospheric emission. 

\subsection{\object{TYC 428-980-1}} 
\object{TYC 428-980-1} displays excesses in $W3$, $W4$, and MIPS 24~$\mu$m. While the MIPS1 PSF fit shows a $\chi^2$ of 1.3, the source is flagged as an extended object in the WISE catalogue. Figures~\ref{fig:images-Spitzer} and \ref{fig:image-WISE} show a close source at a separation of around $12\arcsec$ but $\sim100$~pc closer according to the \textit{Gaia} DR2 parallaxes. Such a separation could be spatially resolved in the WISE $W1-W3$ bands (not in $W4$), and in MIPS 24~$\mu$m. The fact that this close source is detected neither in the $W3$ band nor in the MIPS1 image implies that the detected emission is associated to the central star. The SED of this object (see Fig.~\ref{fig:SED}) shows that the $W4$ (22~$\mu$m) and MIPS1 (24~$\mu$m) excesses are compatible, making an extremely interesting new candidate.

\subsection{\object{2MASS J17462472+0517213}}

This source has an excess in MIPS at 24~$\mu$m. We cannot confirm the excess with WISE since the this source is detected as an upper limit in $W4$. This is the coolest star for which we detect an IR excess. VOSA finds a best fit with a SED of 5\,750~K consistent with the \textit{Gaia} and \citet{Smith+11} effective temperatures and with an early G spectral type.

\subsection{\object{HD 161734}}

HD~161734 was not initially included in our sample because it has a low membership probability in \citet{Miret-Roig+19}. However, \citet{Smith+11} and \citet{Meng+2017} detected an excess emission from the near to mid-infrared which motivated us to study it further. With our new IRAC and MIPS photometric reduction, we also detect an excess. Additionally, we searched the WISE photometry finding an excess consistent with what is seen with \textit{Spitzer} (see Fig.~\ref{fig:SED}). \citet{Smith+11} mention in their conclusions that this source may be a binary. However, they did not mention on what they based their suspects and we did not find any reason to believe so. Based on its near-IR excess, they proposed that this source could have a remnant primordial disc. In fact, they can fit the excess with a 500\,K blackbody, suggesting a dusty disc with a radius of 1.7~AU. Considering the members obtained with \textit{Gaia} in \citet{Miret-Roig+19} the cluster has a median and standard deviation parallax of 2.84~mas and 0.36~mas, respectively. For the proper motions, the median and standard deviation are $-0.91$~mas~yr$^{-1}$ and 0.64~yr$^{-1}$ in right ascension and $-8.49$~mas~yr$^{-1}$ and 0.68~yr$^{-1}$ in declination. The \textit{Gaia} DR2 astrometry of HD\,161734 is $\varpi = 2.1716\pm0.0407$~mas, pmRA~$=-1.623\pm0.063$~mas~yr$^{-1}$, pmDec~$= -9.432\pm0.062$~mas~yr$^{-1}$. Therefore, we see that this source has a very precise \textit{Gaia} DR2 astrometry which falls further than $1\sigma$ of the cluster distribution in all the spaces, specially in parallax which is a decisive variable for the membership analysis. However, we checked that this source has a photometry that agrees well with the main sequence of the cluster, and the fact that it shows a clear IR excess makes it a good debris disc candidate member. Future \textit{Gaia} releases with improved astrometry might rise the membership probability of this source. In any case, the membership of \citet{Miret-Roig+19} has a true positive rate of $\sim90\%$, and this source is an example of the objects that could be missing in that  list of members.

\section{Discussion}
\label{sec:discussion}

\begin{figure}
    \centering
    \includegraphics[width = \columnwidth]{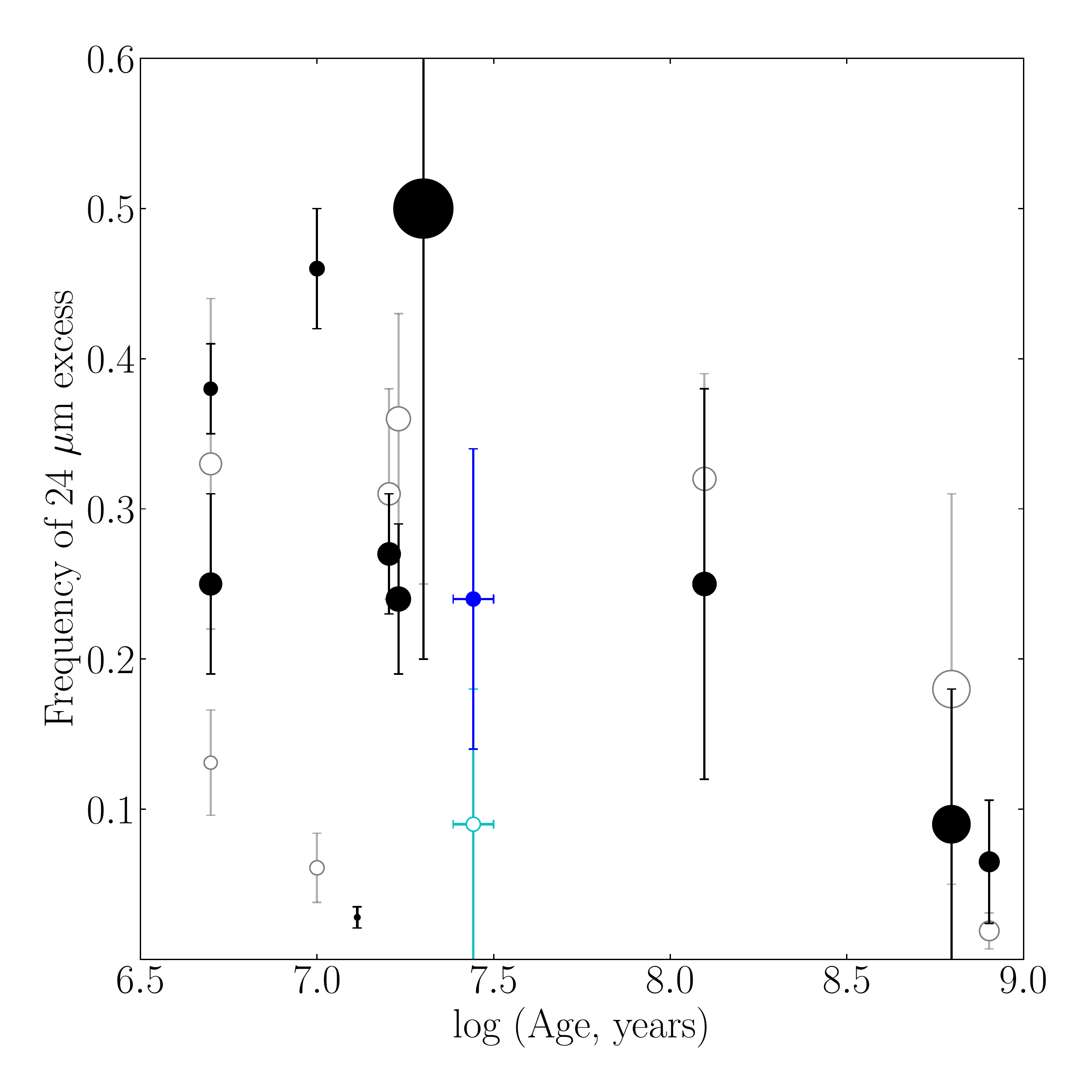}
    \caption{Frequency of 24~$\mu$m excess in the sample listed in Table~1 of \citet{Chen+2020} for early type stars ($2-2.5$~M$_\odot$, black filled dots) and for solar type stars ($1-1.5$~M$_\odot$, grey empty dots). The frequencies measured in this work are marked with a blue filled dot for B--A stars and with a cyan empty dot for solar type stars. All the markers have sizes inversely proportional to the distance of the cluster. }
    \label{fig:disc_evolution}
\end{figure}{}

We have estimated that our MIPS photometry is complete down to [24]~$\lesssim 11.8$ (late-B to mid-K, see Sect.~\ref{subsec:completeness}) in our sample. In this spectral range and in the central $1\degr \times 1\degr$ area covered by MIPS (37~pc$^2$), the disc fraction of IC~4665 is 5/32 or $16\pm7$\%. This fraction is smaller (although compatible within the uncertainties) than the 27$_{-7}^{+9}$\% rate reported in a previous study of this cluster, covering the same field of view and magnitude range \citep{Smith+11}. The main reasons for this difference are our improved image-processing techniques (see Sect.~\ref{sec:data}), the fact that we only provide PSF photometry and they mixed PSF and aperture photometry, and the different lists of members. We discarded most of their candidates, because their MIPS1 photometry was systematically brighter than ours (10 sources were rejected for this reason), and/or they are no longer classified as  members after the analysis of the \textit{Gaia} DR2 astrometry (3 sources were rejected for this reason).

Many studies in the literature provide the disc fraction for B--A stars and solar type stars, separately, in different clusters and star forming regions. To compare our study with these results, we have estimated these fractions in our sample. The B--A stars in IC~4665 have an intrinsic colour $G-Ks\lesssim0.75$~mag and, in this range, the disc fraction becomes 4/17 or $24\pm10$\%. 
If we apply the same selection to the sample of \citet{Smith+11} we obtain a disc fraction of 5/14 or $36\pm12$\% for B--A stars. Therefore, in this spectral range we also obtain a smaller disc fraction, although both are consistent within the uncertainties. 
We only detect one candidate in the spectral range F5--K5 (solar type stars) which results in a disc fraction of 1/11 or $9\pm9$\%. \citet{Smith+11} report a disc fraction of 10/24 or $42_{-13}^{+18}\%$ in the same spectral range which is discrepant with our results. Finally, we note that the disc fractions derived by us for early and solar-type stars are compatible within the relatively large uncertainties. 

In the following, we compare the disc fractions obtained in this study with other young clusters and associations. This comparison should be regarded as tentative and taken with caution, given that the various studies quoted below have very different levels of sensitivity and/or completeness (see \citealt{Wyatt+2008, Hughes+18}, and references therein for a detailed discussion on the difficulties related to such comparisons). Additionally, the level of completeness and contamination in the list of members differs from one study to another and most of them are based on pre-\textit{Gaia} members lists. 

\citet{Gorlova+07} did a study analogous to the one presented here for the NGC~2547 open cluster. This is a very similar cluster in terms of age (30~Myr, \citealt{Jeffries+06}) and distance (400~pc). They imaged the inner $1\degr \times 1\degr$ regions which at the distance of the cluster corresponds to $\sim50$~pc$^2$, and were complete down to a spectral type of late-F in MIPS1. 
They found a B8--A9 excess fraction of $\sim$44\% and a F0--F9 excess fraction of $\sim$ 33\%. These values are significantly larger than what we find in IC~4665, although the authors do not provide uncertainties.
We believe that the same reasons we discussed to explain the differences with the study of \citet{Smith+11} could apply to this discussion. Indeed, we checked that around 40\% of their MIPS sample could be contaminants according to the \textit{Gaia} DR2 astrometry.

\citet{Gorlova+06} studied the disc population of the intermediate age Pleiades cluster (120~Myr). They analysed \textit{Spitzer} MIPS1 data of an area covering the central $2\degr\times1\degr$ area of the cluster which at the distance of the Pleiades corresponds to an area of 14~pc$^2$. They were complete down to a spectral type of K3, or even M2 in the regions with less nebulosity. They estimated the debris disc fraction of B--A members to be $\sim$25\%. The value is very similar to the one found in IC~4665 and is consistent with a slow evolution of the 24~$\mu$m excess in debris discs, with a characteristic timescale of 150\,Myr \citep[e.g.][]{Siegler+07, Gorlova+06}. 
However, the proximity of the Pleiades with respect to IC~4665 leads to a significantly smaller spatial coverage of this cluster, hindering a proper comparison of the disc fractions.

In Figure~\ref{fig:disc_evolution} we compare the disc fractions obtained in this study with several nearby clusters and associations reported in a recent work by  \citet{Chen+2020} for early and solar type stars. We see that our disc fraction for B--A stars is compatible with the disc evolution trend defined by the other clusters. The disc fraction we measure at 30~Myr is compatible within the uncertainties with clusters of 15--20~Myr (Upper Centaurus Lupus, Lower Centaurus Crux, and the $\beta$~Pictoris moving group) and with the Pleiades at 125~Myr. In the case of solar type stars, our disc fraction is smaller than that from clusters of $\sim$20 Myr which are at distances closer than 150~pc. Interestingly, our disc fraction is similar to the one reported in younger clusters (5--10~Myr) at similar distances (Orion OB1a and Orion OB1b). 

Our \textit{Spitzer} photometry of IC\,4665 only covers the central $1\degr\times1\degr$ of the cluster (see Fig.~\ref{fig:Spitzer_coverage}). However, according to the most complete membership analysis to date \citep{Miret-Roig+19}, the cluster has a size of at least 3\degr\ of radius. We estimated that MIPS1 observations only cover 55\% of the B--A stars of the cluster by comparing the spatial distribution of all the B--A members in a circle of 3\degr\ radius (whole cluster area) with the same population in the area covered by MIPS data ($1\degr\times1\degr$). Therefore, the disc fractions we obtain with MIPS1 are in principle only valid for the central part of the cluster. 
Since we also have WISE photometry, which covers all the area occupied by the cluster (a circle of 3\degr\ radius in this case) we used it to to estimate the disc fraction. We obtain a fraction of 3/14 or 29$\pm12$\% at 22~$\mu$m ($W4$) in the spectral range late--B to mid--A (see Sect.~\ref{subsec:completeness}). This value is very similar to what we obtain with MIPS at the central region in the same spectral range 2/8 or $25\pm15$\%.

\section{Conclusions}
\label{sec:conclusions}

In this paper we presented a study of the debris disc population of the 30~Myr open cluster IC~4665. We identified six candidates with IR excess, two of which are new candidates. All of them have excesses at 22--24~$\mu$m but not in the near infrared (except of HD~161734), indicative of the presence of dusty debris discs. Using MIPS data in the central part of the cluster, we computed a disc fraction of 24$\pm$10\% B--A stars. 
We only detected one solar type star resulting in a fraction of $9\pm9\%$ in the F5--K5 spectral type range.
Using WISE, we extended the search to the outskirts of the cluster finding a similar result in the B--A range. We believe that the main differences between our results and the ones obtained in a similar study of this cluster \citep{Smith+11} are mainly due to i) the fact that we used improved image-processing techniques, ii) we only measured PSF photometry and they combined PSF and aperture photometry, and iii) we worked with different member lists.

Comparing our disc fraction with other nearby clusters and associations we found that for early type stars our results are compatible with regions of 15--20~Myr  (Upper Centaurus Lupus, Lower Centaurus Crux the $\beta$~Pictoris moving group), and 125~Myr (the Pleiades). For solar type stars we find a disc fraction lower than generally observed at 15--20~~Myr. 
We would like to emphasise that all the studies we used to compare are pre-\textit{Gaia}, meaning that the derived  memberships (and in consequence disc fractions) should be revised with the new astrometry. Additionally, we have seen the importance of the image processing, specially at 24~$\mu$m, where the images are strongly affected by nebulosities.

All our candidates were first detected in colour-colour diagrams and using an empirical photospheric sequence to define the non-excess locus. We also used the Kurucz atmospheric models to fit a SED and independently detect IR excess. As an extra product of this procedure, we obtained effective temperatures which are more accurate than the photometric ones available in the literature. Our candidates are good targets to be followed up with ALMA and with the future mission James Webb Space Telescope (JWST).

\appendix

\section{Additional Tables and Figures}

\begin{figure}
    \centering
    \includegraphics[width = \columnwidth]{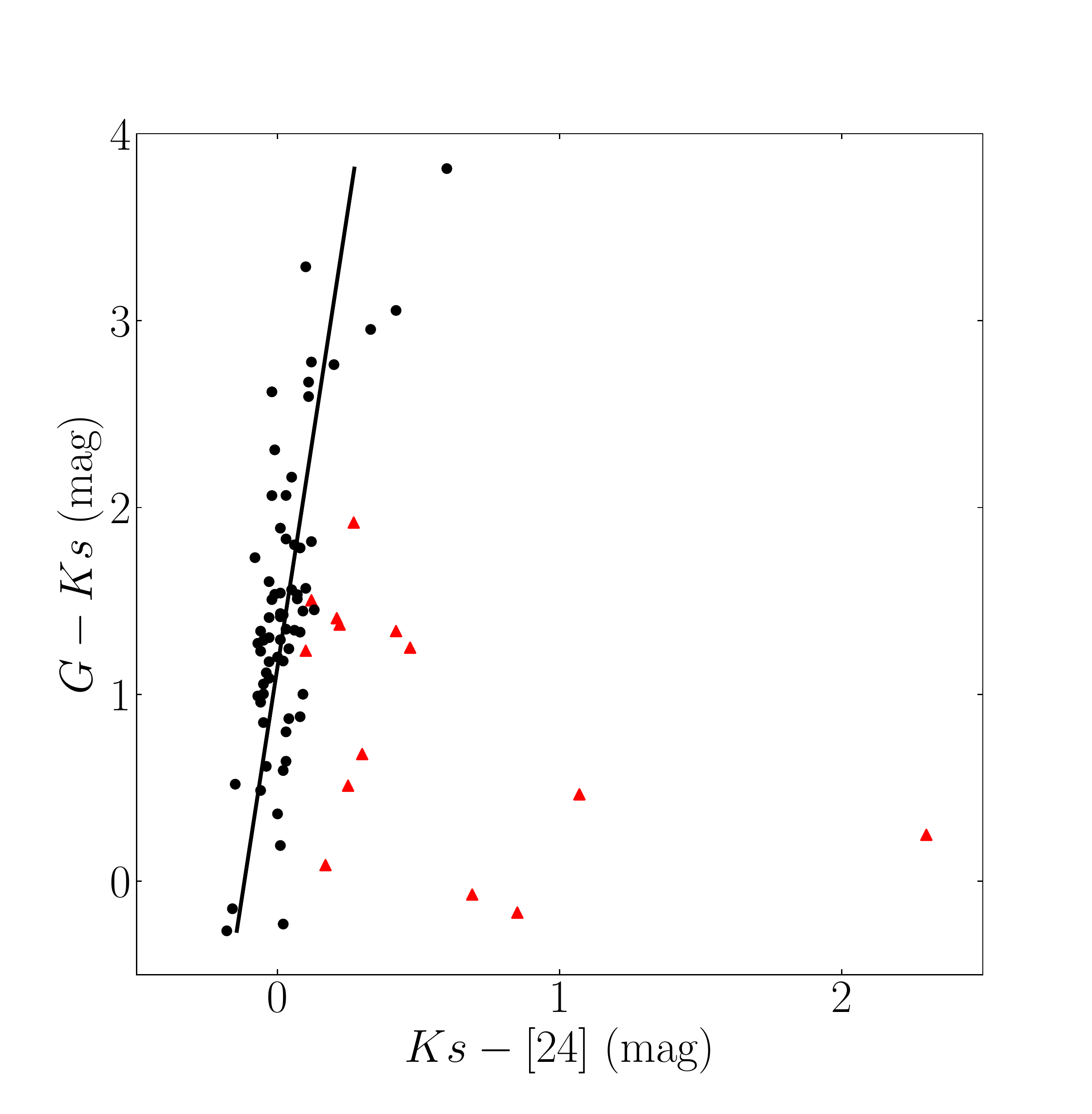}
    \caption{$(G-Ks)-(Ks-[24])$ colour-colour diagram of the Pleiades members with good 24~$\mu$m photometry from Table~2 of \citet{Gorlova+06}. Only the sources classified as non-excess sources by the authors (black points) are used in the fit. } 
    \label{fig:gorlova+06}
\end{figure}{}

\begin{figure}
    \centering
    \includegraphics[width = \columnwidth]{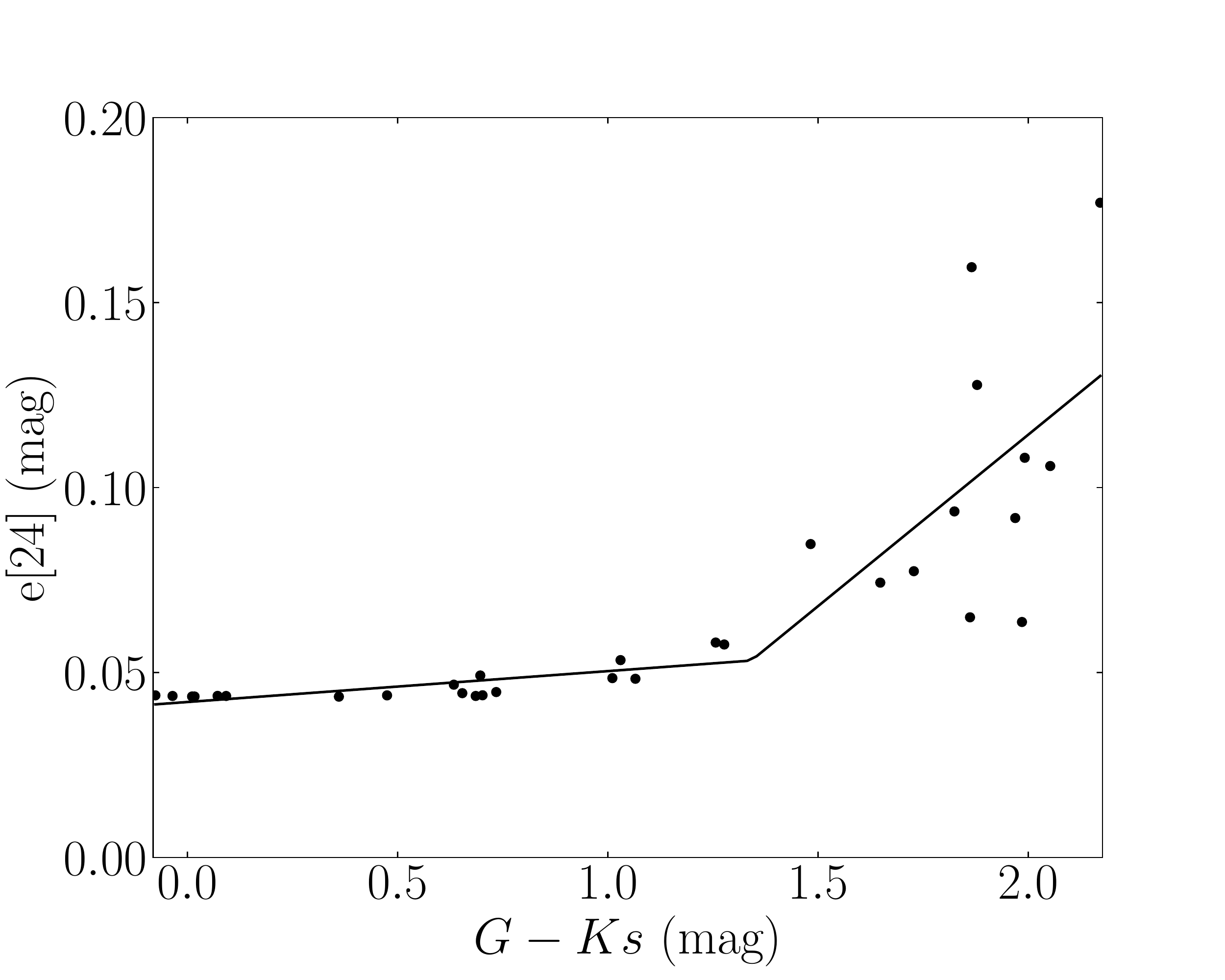}
    \caption{Photometric uncertainties at [24]~$\mu$m as a function of the $G-Ks$ colour.}
    \label{fig:fit_eMIPS1}
\end{figure}{}

\begin{figure}
    \centering
    \includegraphics[width = \columnwidth]{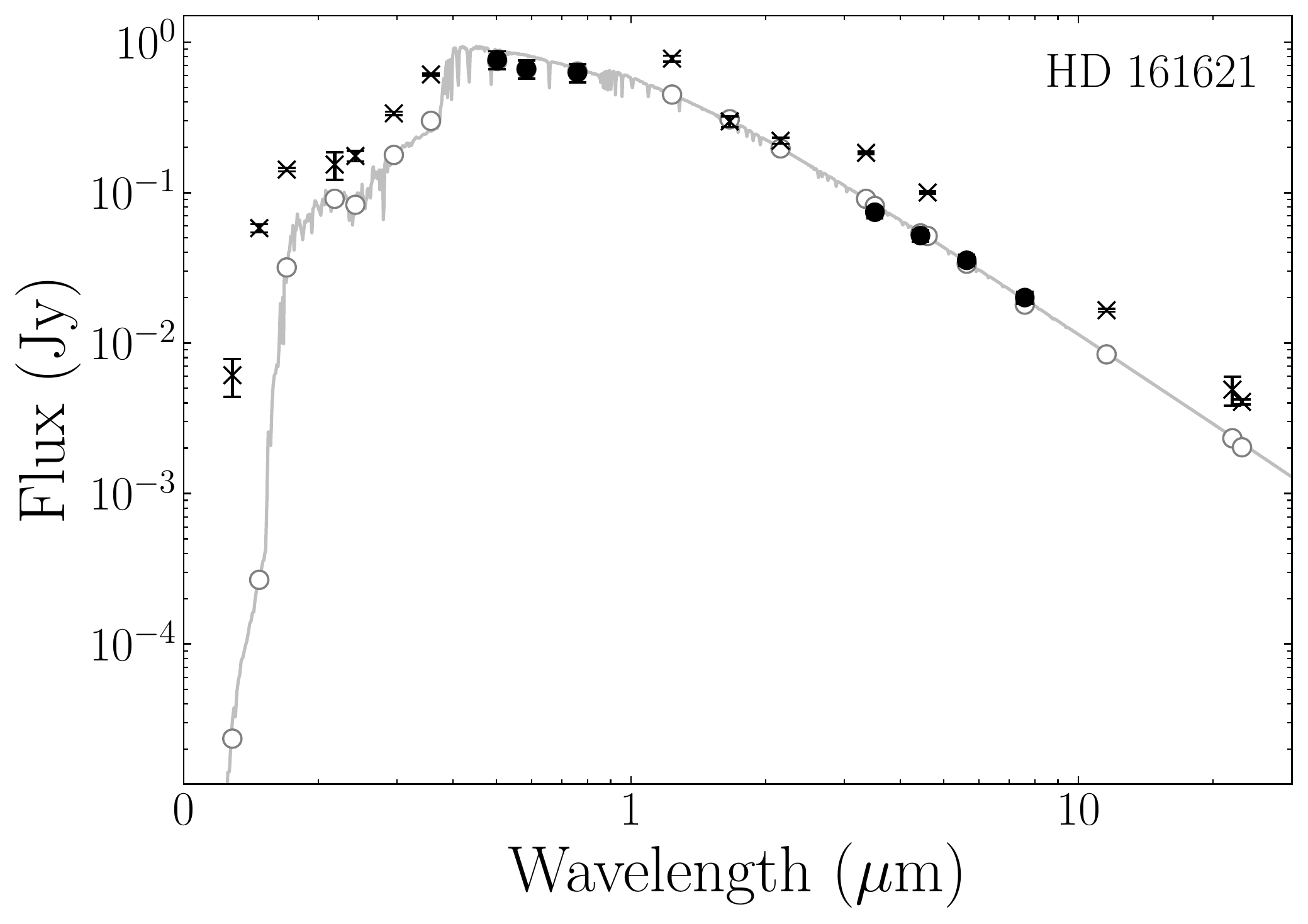}
    \caption{Same as Figure~\ref{fig:SED} for the source \object{HD 161261} which does not present an IR excess. The filled black circles represent the photometric points of the primary, since in these observations the binary system is resolved.}
    \label{fig:SED-387}
\end{figure}{}

\begin{landscape}
\setlength{\tabcolsep}{3pt} 
\renewcommand{\arraystretch}{1.5} 
\begin{table}
\begin{center}
\caption{Photometric bands used to detect an IR excess in Sect.~\ref{sec:colour-colour-diagrams}. Only the sources for which we have detected an IR excess are shown. An extended version of this Table including all the photometric bands analysed in this work for all the members of IC~4665 is available at the CDS.}
\label{tab:photometry}
\begin{tabular}{|l|r|r|r@{$\pm$}l|r@{$\pm$}l|r@{$\pm$}l|r@{$\pm$}l|r|r|r|r|r|}
\hline
\hline
  \multicolumn{1}{|c|}{Object} &
  \multicolumn{1}{c|}{$G$} &
  \multicolumn{1}{c|}{$Ks$} &
  \multicolumn{2}{c|}{$W1$} &
  \multicolumn{2}{c|}{$W2$} &
  \multicolumn{2}{c|}{$W3$} &
  \multicolumn{2}{c|}{$W4$} &
  \multicolumn{1}{c|}{IRAC1} &
  \multicolumn{1}{c|}{IRAC2} &
  \multicolumn{1}{c|}{IRAC3} &
  \multicolumn{1}{c|}{IRAC4} &
  \multicolumn{1}{c|}{MIPS1} \\

  \multicolumn{1}{|c|}{ } &
  \multicolumn{1}{c|}{(mag)} &
  \multicolumn{1}{c|}{(mag)} &
  \multicolumn{2}{c|}{(mag)} &
  \multicolumn{2}{c|}{(mag)} &
  \multicolumn{2}{c|}{(mag)} &
  \multicolumn{2}{c|}{(mag)} &
  \multicolumn{1}{c|}{(mag)} &
  \multicolumn{1}{c|}{(mag)} &
  \multicolumn{1}{c|}{(mag)} &
  \multicolumn{1}{c|}{(mag)} &
  \multicolumn{1}{c|}{(mag)} \\
\hline
     \object{HD 161261} &  8.26$\pm$0.05 & 8.04$\pm$0.05 & 8.00 & 0.02$^{*}$ & 7.98 & 0.02       & 7.75 & 0.02       & 7.15 & 0.11                     &               &               &               &               &  \\
     \object{HD 161733} &  7.96$\pm$0.05 & 7.87$\pm$0.05 & 7.85 & 0.03$^{*}$ & 7.89 & 0.02$^{*}$ & 7.87 & 0.02       & 7.69 & 0.15                     & 7.92$\pm$0.03 & 7.92$\pm$0.03 & 7.99$\pm$0.03 & 7.92$\pm$0.03 & 7.75$\pm$0.04\\
     \object{HD 161621} &  9.45$\pm$0.05 & 8.75$\pm$0.05 & 8.09 & 0.02$^{*}$ & 8.10 & 0.02$^{*}$ & 8.14 & 0.02$^{*}$ & 8.08 & 0.24$^{*}$               & 8.96$\pm$0.03 & 8.87$\pm$0.03 & 8.78$\pm$0.03 & 8.74$\pm$0.03 & 8.11$\pm$0.04\\
\object{TYC 428-1938-1} & 11.01$\pm$0.05 & 9.95$\pm$0.05 & 9.85 & 0.02       & 9.86 & 0.02       & 9.80 & 0.05       & \multicolumn{2}{l|}{$>$8.64} & 9.87$\pm$0.03 & 9.83$\pm$0.03 & 9.80$\pm$0.03 & 9.84$\pm$0.03 & 9.46$\pm$0.05\\
 \object{TYC 428-980-1} & 10.36$\pm$0.05 & 9.68$\pm$0.05 & 9.40 & 0.02$^{*}$ & 9.53 & 0.02$^{*}$ & 9.36 & 0.04$^{*}$ & 7.72 & 0.18$^{*}$               & 9.62$\pm$0.03 & 9.64$\pm$0.03 & 9.60$\pm$0.03 & 9.58$\pm$0.03 & 7.77$\pm$0.04\\
\object{2MASS J17462472+0517213}  & 12.67$\pm$0.05 & 10.81$\pm$0.05 & 10.81 & 0.02 & 10.66 & 0.02 & 10.56 & 0.09 & \multicolumn{2}{l|}{$>$8.97} & 10.69$\pm$0.03 & 10.80$\pm$0.03 & 10.65$\pm$0.03 & 10.67$\pm$0.05 &10.32$\pm$0.06\\
     \object{HD 161734} &  8.82$\pm$0.05 & 8.46$\pm$0.05 & 8.34 & 0.02       & 8.25 & 0.02       & 7.64 & 0.02       & 6.75 & 0.07                     & 8.40$\pm$0.03 & 8.31$\pm$0.03 & 8.13$\pm$0.03 & 7.86$\pm$0.03 & 6.99$\pm$0.04\\
\hline
\hline\end{tabular}
    
\end{center}{}
\noindent\tablefoottext{*}{measurement filtered (see Sect.~\ref{subsec:phot_filter})}\\

\end{table}
\end{landscape}

\begin{acknowledgements}
We acknowledge the referee for the exhaustive revision of our work which improved the quality of our study.
We thank LM Sarro for his comments on the \textit{Gaia} effective temperatures and the membership probability of HD~161734.
This research has received funding from the European Research Council (ERC) under the European Union’s Horizon 2020 research and innovation programme (grant agreement No 682903, P.I. H. Bouy), and from the French State in the framework of the ”Investments for the future” Program, IdEx Bordeaux, reference ANR-10-IDEX-03-02.
NH has been partially funded by the Spanish State Research Agency (AEI) Project No. ESP2017-87676-C5-1-R and No. MDM-2017-0737 Unidad de Excelencia “María de Maeztu”- Centro de Astrobiología (INTA-CSIC).
This work is based [in part] on observations made with the Spitzer Space Telescope, which is operated by the Jet Propulsion Laboratory, California Institute of Technology under a contract with NASA.
This publication makes use of data products from the Two Micron All Sky Survey, which is a joint project of the University of Massachusetts and the Infrared Processing and Analysis Center/California Institute of Technology, funded by the National Aeronautics and Space Administration and the National Science Foundation.
This publication makes use of data products from the Wide-field Infrared Survey Explorer, which is a joint project of the University of California, Los Angeles, and the Jet Propulsion Laboratory/California Institute of Technology, funded by the National Aeronautics and Space Administration. 
This research has made use of the NASA/ IPAC Infrared Science Archive, which is operated by the Jet Propulsion Laboratory, California Institute of Technology, under contract with the National Aeronautics and Space Administration.
This publication makes use of VOSA, developed under the Spanish Virtual Observatory project supported by the Spanish MINECO through grant AyA2017-84089.  
This work made extensive use of the following softwares and libraries: Parallel \citep{Tange2011a}, astropy \citep{astropy:2013}, Topcat \citep{2005ASPC..347...29T}.
This research has made use of the VizieR catalogue access tool, CDS, Strasbourg, France. The original description of the VizieR service was
 published in A\&AS 143, 23.
 \end{acknowledgements}


\bibliographystyle{aa} 
\bibliography{mybiblio.bib}

\end{document}